\documentstyle[12pt]{article}
\begin{document}
\begin{flushright}
hep-th/0205135\\
SNBNCBS-2002
\end{flushright}
\vskip 2.5cm
\begin{center}
{\bf \Large { Cohomological aspects of gauge theories: superfield formalism}}

\vskip 3cm

{\bf R.P.Malik}
\footnote{ E-mail address: malik@boson.bose.res.in  }\\
{\it S. N. Bose National Centre for Basic Sciences,} \\
{\it Block-JD, Sector-III, Salt Lake, Calcutta-700 098, India} \\

\vskip 3cm

\end{center}

\noindent
{\bf Abstract}:
Some of the key cohomological features of the two $(1+1)$-dimensional
(2D) free Abelian- and self-interacting non-Abelian gauge theories
(having no interaction with matter fields) are 
briefly discussed first in the language of symmetry properties of the 
Lagrangian densities and the same issues are subsequently addressed 
in the framework of superfield formulation on the four 
($2 + 2)$-dimensional supermanifold. Special emphasis is laid on the 
on-shell- and off-shell nilpotent (co-)BRST symmetries 
that emerge after the application of (dual) horizontality
conditions on the supermanifold. The (anti-)chiral superfields play a very
decisive role in the derivation of the on-shell nilpotent symmetries. The
study of the present superfield formulation leads to 
the derivation of some new symmetries
for the Lagrangian density and the symmetric energy-momentum tensor.
The topological nature of the
above theories is captured in the framework of superfield formulation and
the geometrical interpretations are provided for some of the topologically
interesting quantities.\\

\vskip 0.5cm



$~~~~~$(anti-)chiral superfields; topological
properties; 2D free Abelian- and self-interacting 

$~~~~~$non-Abelian gauge theories

\baselineskip=16pt


\newpage

\noindent
{\bf 1 Introduction}\\

\noindent
The principle of local gauge invariance has played a notable role in the
modern developments of theoretical high energy physics up to the energy
scale of the order of grand unification. The theories, based on such a
principle, are known as gauge theories and they are
endowed with first-class constraints in the language
of Dirac's classification scheme [1,2]. The Lagrangian density of such a class 
of theories is always singular and respects the ``classical''
local gauge symmetry transformations that are generated by these first-class
constraints. In addition to their aesthetic
theoretical appeal, these theories (in particular, one-form gauge theories)
have also shed light on the results of some of the landmark experiments in 
the recent past. One of the most elegant methods for
the covariant canonical quantization of such type of gauge theories is the
Becchi-Rouet-Stora-Tyutin (BRST) formalism which, in its own theoretical
setting, maintains unitarity and ``quantum'' gauge invariance together
at any arbitrary order of perturbation theory. 
In this formalism, the ``classical'' local symmetries of the gauge theories
are traded with the supersymmetric type nilpotent ``quantum'' gauge symmetries
that are usually called the BRST symmetries. In fact, the latter 
symmetries are generated
by a conserved ($\dot Q_{b} = 0$) and nilpotent ($Q_{b}^2 = 0$) BRST charge
$Q_{b}$. The physical states of the theory belong to a {\it subspace} of
the total Hilbert space of states where the physicality condition
$Q_{b} |phys> = 0$ is satisfied. This condition implies the annihilation of 
the physical states of the ``quantum'' gauge theory by the operator form of 
the first-class constraints of the original ``classical'' gauge theory.
Thus, BRST closed $Q_{b} |phys> = 0$
states turn out to be consistent with the Dirac's 
prescription  for the quantization of systems with constraints [1,2]. 
This formalism, in the present scenario of 
the frontier areas of research in theoretical physics,
is indispensable in the context of modern developments in topological field 
theories (TFTs) [3-5], topological superstring theories [6,7] and, in general,
in the domain of (super)string theories, M-theory and D-branes, etc., 
(see, e.g., Ref. [8] and references therein). The range and scope of BRST 
formalism have been beautifully extended to include the second-class 
constraints in its domain of applicability  [9,10]. Its mathematically
elegant inclusion in the well-known Batalin-Vilkovsky formalism [10,11],
its deep connection with the mathematics of differential geometry and 
cohomology [12-14], its clear and transparent geometrical interpretation 
in the framework of superfield
formalism [15-19], etc., have elevated this subject of investigation to a 
fairly high degree of physical as well as mathematical sophistication.

The nilpotency of the BRST charge and physicality condition are the two key
properties that are deeply connected with the cohomological properties of
the closed (i.e. $ d f = 0$) differential forms $f$ w.r.t. the nilpotent
$d^2 = 0$ exterior derivative $d$ (with $d = dx^\mu \partial_\mu$). In fact,
two closed ($d f = d f^\prime = 0$)
forms $f$ and $f^\prime$ are said to belong to the same cohomology
class w.r.t. $d$ if they differ by an exact form (i.e. $f^\prime = d + d g$).
Similarly, two physical states belong to the same cohomology class w.r.t.
the BRST charge $Q_{b}$ if they differ by a BRST exact state. Thus, it is
evident that the operator $d$ of differential geometry finds its analogue
in the BRST charge $Q_{b}$
that generates a local, covariant, continuous and nilpotent
symmetry for the Lagrangian density of a given gauge theory. In addition to 
$d$, there are two
more operators $\delta = \pm * d *$ and $\Delta = (d + \delta)^2$
that form a set $(d, \delta,\Delta)$ of the de Rham cohomology operators of
differential geometry
\footnote{ The operators $\delta$ and $\Delta$ are called the co-exterior 
derivative and Laplacian operators respectively and $*$ is the
Hodge duality operation on the manifold. Together, all these
three operators obey an algebra: $ d^2 = \delta^2 = 0, \Delta = \{ d, \delta\}
= d \delta + \delta d, [\Delta, d ] = [\Delta, \delta] = 0$
implying that $\Delta$ is the Casimir 
operator [20-23].}. In terms of these operators, 
the celebrated Hodge decomposition theorem (HDT) is defined which states that,
on a compact manifold without a boundary, any arbitrary form $f_{n} $ of 
degree $n$ $(n = 0, 1, 2, 3.....)$
can be uniquely written as the sum of a harmonic form $h_n$
($\Delta h_{n} = d h_{n} = \delta h_{n} = 0$), an exact form 
($d e_{n-1}$) and a co-exact form ($\delta c_{n+1}$) as given below [20-23]
$$
\begin{array}{lcl}
f_{n} = h_{n} + d\; e_{n-1} + \delta\; c_{n+1}.
\end{array} \eqno(1.1)
$$
It has been a long-standing problem to find (i) the analogues of $\delta$,
$\Delta$  in the language of symmetry properties 
(and corresponding generators) of the Lagrangian density
of a given gauge theory in any arbitrary spacetime dimension. (ii) The
clear description of and about the cohomology of quantum states in the total
Hilbert space of states w.r.t. the analogues of $d,\delta,\Delta$. 
(iii) The analogue of HDT (cf. (1.1)) in the
language of states of the quantum theory in the total quantum Hilbert space of 
states. Some interesting attempts were made towards this goal for the
interacting (non-)Abelian
gauge theories in any arbitrary spacetime dimension but the relevant symmetry
transformations turned out to be non-local and non-covariant [24-27]. In
the covariant formulation, the nilpotency was restored only for a certain 
specific value of the parameter of these theories [28].

Recently, in a set of papers [29-33], a possible connection between
the local, continuous and covariant symmetries and their generators on
the one hand and the cohomological  operators $(d, \delta, \Delta)$ 
on the other hand, has been established in the Lagrangian formulation for
the case of the 2D free- as well as interacting (non-)Abelian gauge theories.
A discrete symmetry transformation 
has been shown to correspond to the Hodge duality
$*$ operation. The existence of such type of local, covariant and continuous
symmetries as well as a discrete symmetry
has also been shown for the physical $(3+1)$-dimensional (4D)
free 2-form Abelian gauge theory [34]. 
All the above examples provide a beautiful set of tractable 
field theoretical models for the Hodge theory
(from the point of view of mathematics as well as physics).
One of the physical consequences of these
studies has been to establish the fact that the  free 2D Abelian-
and self-interacting non-Abelian gauge theories belong to a new class of TFTs
in the flat spacetime
(i.e., {\it a field theory endowed with a flat spacetime metric and having no
propagating degrees of freedom associated with  the basic field of
the theory}) (see, e.g., [5] for details). In fact, these new TFTs
capture together some of the key topological properties of the
Witten- and Schwarz type TFTs in flat spacetime.
For instance, the appearance of the Lagrangian
densities turns out to be like Witten type TFTs 
(because they are equal to the sum of a BRST- and co-BRST
anti-commutators) but the local symmetries
of these theories are that of the Schwarz type [35]
(as there are no topological shift symmetries in the theory whose
existence is one of the characteristic features of the Witten type TFTs). 
Topological invariants for these theories have been computed and their 
recursion relations have been obtained [29,30,35] on a flat 2D manifold
(which is equivalent to a 2D closed Riemann surface in its Euclidean version).

A different but interesting aspect of the above set
of problems (cited after (1.1)) is to provide the
geometrical origin and interpretation for the conserved and nilpotent
(co-)BRST charges (and the symmetry transformations they generate) in the
language of {\it translations} along the Grassmannian directions of the 
$(2+2)$-dimensional compact supermanifold. Generally,
in the superfield approach [15-19] to BRST formalism for the 
$p$-form ($ p = 1, 2, 3.....$) gauge theories, the curvature ($(p+1)$-form)
tensor is restricted to be flat along the Grassmannian directions of the
$(D+2)$-dimensional supermanifold, parametrized by $D$-number of spacetime
(even) coordinates and two Grassmannian (odd) coordinates.
This flatness condition, popularly known as horizontality condition
\footnote{ This restriction has been referred to as the ``soul-flatness''
condition in Ref. [36] which amounts to setting the Grassmannian components
of the super curvature ($(p+1)$-form) tensor equal to zero.}, provides the 
origin for the existence of (anti-)BRST symmetry transformations and leads to 
the geometrical interpretation of the conserved and nilpotent 
($Q_{(a)b}^2 = 0$) (anti-)BRST charges ($Q_{(a)b}$) as the translation
generators along the Grassmannian directions. 
In this derivation, the super exterior derivative $\tilde d$
and the Maurer-Cartan equation  (for the definition of the curvature
tensor) are exploited together for the imposition of the 
horizontality condition. Recently, in a couple of papers [37,38], 
all the three super de Rham cohomology operators 
($\tilde d, \tilde \delta, \tilde \Delta$),
defined on the $(2+2)$-dimensional supermanifold, have been exploited to
show the existence of nilpotent (anti-)BRST- and (anti-)co-BRST 
symmetries as well as a bosonic symmetry (which is equivalent to the 
anti-commutators of the nilpotent (anti-)BRST and (anti-)co-BRST symmetries)
in the framework of superfield formulation for the case of a free 
2D Abelian gauge theory. 
Such kind of geometrical superfield formulation has also been carried out for
the self-interacting 2D non-Abelian gauge theory [39]. In these discussions
and derivations, a generalized version of the horizontality conditions w.r.t.
the super cohomological operators $(\tilde d, \tilde \delta, \tilde \Delta)$
has been used. The topological nature
of these theories has also been captured in the (chiral) superfield formulation
and the geometrical interpretations for some of the physically interesting
quantities have been briefly discussed [40,41]. In the relevant 
and pertinent literature available on the subject under discussion,
there are many ways to define the topological nature of a field theory. 
However, it should be re-emphasized at this stage that, in all our 
earlier works (e.g. [29],[30],[35],[40],[41]), we have
taken the point of view that a theory is topological in a given flat spacetime
if there are no propagating degrees of freedom associated with the basic
field(s) of the theory (see. e.g., [5] for details).

The purpose of the present paper is to capture the on-shell and off-shell
nilpotent symmetries, Lagrangian density, symmetric energy momentum tensor,
topological invariants, etc., for the 2D free Abelian- and self-interacting 
non-Abelian gauge theories in the language of superfield formulation on
a  four $(2+2)$-dimensional supermanifold. It is essential and
important to lay emphasis
on the fact that, to the best of our knowledge, the on-shell nilpotent
symmetries for the (non-)Abelian gauge theories have not yet been discussed
and derived in the framework of superfield approach to BRST formalism
despite a rich literature available on this subject
connected with the one-form and 2-form gauge theories. In our present 
paper, for the discussion of the
on-shell nilpotent symmetries, we invoke the (anti-)chiral superfields in terms 
of which the corresponding Lagrangian density, symmetric energy momentum tensor
and topological invariants are expressed. This exercise leads to the 
geometrical interpretation for the above physically interesting quantities
in the language of translations along some {\it specific} direction of the 
supermanifold. For the 2D free Abelian gauge theory, the choice of the 
(anti-)chiral superfields enables us to demonstrate that the 
on-shell nilpotent (anti-)BRST
and (anti-)co-BRST symmetries co-exist together for a given Lagrangian density
(see, e.g., eqns. (2.1) and (2.2) below). However, for the self-interacting 
2D non-Abelian gauge theory, it turns out that only the {\it on-shell} 
nilpotent (co-)BRST symmetries exist for a given Lagrangian density when
we choose chiral superfield for the description of this theory on a
$(2+2)$-dimensional supermanifold. The {\it on-shell nilpotent} anti-BRST and
anti-co-BRST symmetries do not exist for the same Lagrangian density if we 
choose the anti-chiral superfield on the same supermanifold. This feature
is drastically different from the free 2D Abelian gauge theory. {\it In fact,
with the best of our familiarity with the relevant 
literature, the on-shell nilpotent anti-BRST and
anti-co-BRST symmetries do not exist for the (self-)interacting
non-Abelian gauge theories in any spacetime dimension.} We provide, in our
present paper, an
explanation for this discrepancy in the language of superfield formulation on 
the $(2+2)$-dimensional supermanifold. These on-shell nilpotent (anti-)BRST 
and (anti-)co-BRST symmetries are exploited to provide the geometrical 
interpretation for some of the key topological properties of 
the free 2D Abelian- and 
self-interacting non-Abelian gauge theories. For the discussion of the
off-shell nilpotent (anti-)BRST and (anti-)co-BRST symmetries, we choose the
most general superfields ({\it not} the (anti-)chiral)  on the above 
supermanifold and show that the (dual) horizontality conditions w.r.t.
super cohomological operators $\tilde d$ and $\tilde \delta$, lead to
the geometrical interpretation for the corresponding conserved and nilpotent 
charges as translation generators along the Grassmannian directions of
the supermanifold.

Our present study is essential on three counts. First,
to the best of our knowledge, the super co-exterior derivative $\tilde \delta$
has not yet been exploited extensively in the context of BRST formalism (except
in our recent works [37-41]). Thus, it is an interesting
endeavour to tap the full potential of this operator in as many distinct and
diverse ways as possible and to understand the 
geometry behind it in the framework of superfield formulation. 
In our present paper, for instance, we show that the nilpotent (anti-)co-BRST
symmetries and some of the topologically interesting quantities owe their 
origin to this operator in the framework of superfield approach to
BRST formalism. Second, the insights gained in the context of
2D one-form gauge theories might turn out to be quite useful for similar
discussions in the context of 4D 2-form gauge theories where the existence 
of (anti-)co-BRST and (anti-)BRST symmetries has already been shown [34].
In fact, in a recent work [42], we have been able to show the 
quasi-topological (see, e.g., Ref. [5] for details) nature of the
free 2-form Abelian  gauge theory
in the Lagrangian formulation where the cohomological properties and 
quasi-topological nature are found to be very elegantly intertwined. 
Finally, the geometrical understanding 
of some of the topological features of 2D theories {\it might}
 play an important role in the context of discussion about the
(super)string theories and topological 2D gravity 
where normally a non-trivial (spacetime dependent) 
metric is taken into account for the study of gauge theories in
curved spacetime background.

The contents of our present paper are organized as follows. In section 2,
we set up the notations (as well as conventions)
and briefly recapitulate the essentials of the on-shell
and off-shell nilpotent (anti-)BRST and (anti-)co-BRST symmetries in the 
Lagrangian formulation for the 2D free Abelian gauge theory. These symmetries
are subsequently derived in the framework of superfield formulation by
exploiting the (dual) horizontality conditions w.r.t. the
super cohomological operators $\tilde \delta$ and $\tilde d$. The choice
of (anti-)chiral superfields 
is shown to help in the derivation of on-shell nilpotent symmetries.
Section 3 is devoted to the derivation of on-shell nilpotent (co-)BRST
symmetries and off-shell nilpotent (anti-)BRST and (anti-)co-BRST symmetries
for the self-interacting 2D non-Abelian gauge theory in the framework of
Lagrangian- and superfield formulations. Section 4 deals with
the discussion of topological aspects of the above
theories in the superfield formulation. Finally,
our paper ends with a few conclusions along with some future perspectives
in section 5.\\

\noindent
{\bf 2 Free 2D Abelian gauge theory} \\

\noindent
We discuss here the on-shell- and off-shell nilpotent (anti-)BRST
and (anti-)co-BRST
symmetries for a 2D Abelian gauge theory in the Lagrangian and
superfield formulations.\\

\noindent
{\bf 2.1 (Anti-)BRST- and (anti-)co-BRST symmetries: Lagrangian formulation}\\

\noindent
Let us start off with the BRST invariant Lagrangian density ${\cal L}_{b}$
[43-46] for a non-interacting two ($1 + 1)$-dimensional Abelian gauge theory in 
any arbitrary gauge with a parameter $\xi$
$$
\begin{array}{lcl}
{\cal L}_{b} = - \frac{1}{4}\; F^{\mu\nu} F_{\mu\nu}
- \frac{1}{2 \xi}\; (\partial_{\mu} A^{\mu})^2
- i \;\partial_{\mu} \bar C \partial^\mu C 
\equiv \frac{1}{2}\; E^2 
- \frac{1}{2 \xi} \; (\partial_{\mu}  A^{\mu})^2
- i \;\partial_{\mu} \bar C  \partial^\mu C, 
\end{array} \eqno(2.1)
$$
where $F_{\mu\nu} = \partial_\mu A_\nu - \partial_\nu A_\mu$ 
(with $F_{01} = E$) is the field strength tensor constructed from the
vector potential $A_\mu$, $\xi \neq 0$  and the (anti-)ghost fields $(\bar C)C$
(with $\bar C^2 = C^2 = 0, \; C \bar C + \bar C C = 0)$ are required in the 
theory to maintain the unitarity and ``quantum'' gauge invariance together 
at any arbitrary order of perturbation theory
\footnote{We follow here the conventions and notations such that the 2D flat
Minkowski metric is: $\eta_{\mu\nu} =$ diag $(+1, -1)$ and $\Box = 
\eta^{\mu\nu} \partial_{\mu} \partial_{\nu} = (\partial_{0})^2 - 
(\partial_{1})^2, \varepsilon_{\mu\nu} = - \varepsilon^{\mu\nu}, 
\varepsilon_{01} = + 1, F_{01}
= F^{10} = E = - \varepsilon^{\mu\nu} \partial_\mu A_\nu$. Here the Greek 
indices: $\mu, \nu, \rho...= 0, 1$ correspond to the spacetime directions
on the 2D ordinary manifold.}. 
The above Lagrangian density respects the following on-shell
($\Box C = \Box \bar C = 0$) nilpotent ($s_{(a)b}^2 = s_{(a)d}^2 = 0$)
transformations
$$
\begin{array}{lcl}
s_{b} A_{\mu} &=& \partial_{\mu} C, \qquad s_{b} C = 0, \qquad 
s_{b} \bar C = - \frac{i}{\xi} (\partial_\mu A^\mu), \qquad
s_{ab} A_{\mu} = \partial_{\mu} \bar C, \nonumber\\
 s_{ab} \bar C &=& 0, \qquad 
s_{ab} C = + \frac{i}{\xi} (\partial_\mu A^\mu), \qquad
s_{d} A_{\mu} = - \varepsilon_{\mu\nu} \partial^\nu \bar C, \qquad
s_{d} \bar C = 0, \nonumber\\
 s_{d} C &=& - i E,\qquad 
s_{ad} A_{\mu} = - \varepsilon_{\mu\nu} \partial^\nu C, \qquad
s_{ad} C = 0, \qquad s_{ad} \bar C = + i E,
\end{array}\eqno(2.2)
$$
where $s_{(a)b}$ and $s_{(a)d}$ stand for the (anti-)BRST and (anti-)co-BRST
symmetry operations on the basic fields (with $s_{b} s_{ab} + s_{ab} s_{b} = 0$
and $s_{d} s_{ad} + s_{ad} s_{d} = 0$)
\footnote { Here the conventions and notations of Ref. [46] have been adopted.
The (co-)BRST transformations $\delta_{(D)B}$ (with $\delta_{(D)B}^2 = 0$), 
in their totality, are equivalent to the product of an anti-commuting 
($\eta C = - C \eta, \eta \bar C = - \bar C \eta$) parameter $\eta$ and the
transformations $s_{(d)b}$ (i.e. $\delta_{(D)B} = \eta\; s_{(d)b}$)
where $s_{(d)b}^2 = 0$.}. One can linearize the kinetic energy and
the gauge-fixing terms of the Lagrangian density (2.1) by invoking a
couple of auxiliary fields ${\cal B}$ and $B$. The ensuing Lagrangian
density [29,30,35]
$$
\begin{array}{lcl}
{\cal L}_{B} =  {\cal B}  E - \frac{1}{2} {\cal B}^2 
+ B (\partial_{\mu}  A^{\mu}) + \frac{1}{2} \;\xi\;B^2
- i \partial_{\mu} \bar C \partial^\mu C, 
\end{array} \eqno(2.3)
$$
respects the off-shell nilpotent version of symmetries (2.2) 
as given below [35]
$$
\begin{array}{lcl}
\tilde s_{b} A_{\mu} &=& \partial_{\mu} C, \quad \tilde s_{b} C = 0, \quad 
\tilde s_{b} \bar C = i B, \quad \tilde s_{b} B = 0, \quad 
\tilde s_{b} {\cal B} = 0, \nonumber\\
\tilde s_{ab} A_{\mu} &=& \partial_{\mu} \bar C, \quad
\tilde  s_{ab} \bar C = 0, \quad \tilde s_{ab}  C = - i B, \quad 
\tilde s_{ab} B = 0, 
\quad \tilde s_{ab} {\cal B} = 0, \nonumber\\
\tilde s_{d} A_{\mu} &=& - \varepsilon_{\mu\nu} \partial^\nu \bar C, \quad
\tilde s_{d} \bar C = 0, \quad
\tilde s_{d} C = - i {\cal B},\quad  
\tilde s_{d} {\cal B} = 0, \quad \tilde s_{d} B = 0,
\nonumber\\
\tilde s_{ad} A_{\mu} &=& - \varepsilon_{\mu\nu} \partial^\nu C, \quad
\tilde s_{ad} C = 0, \quad 
\tilde s_{ad} \bar C = + i {\cal B}, \quad \tilde s_{ad} {\cal B} = 0,
\quad \tilde s_{ad} B = 0.
\end{array}\eqno(2.4)
$$
The Lagrangian densities (2.1) and (2.3) respect two more continuous
and covariant symmetries. The anti-commutator of the two nilpotent symmetries
\footnote{ We use here, and in what follows until (2.7), only the notations
for the on-shell nilpotent symmetries but our statements are valid for the 
off-shell nilpotent symmetries as well.} 
$s_{w} = \{ s_{b}, s_{d} \} = \{ s_{ab}, s_{ad} \}$ leads to the existence
of a bosonic (i.e. $s_{w}^2 \neq 0$) symmetry transformation $s_{w}$. The other
symmetry is the ghost symmetry transformation $s_{g}$ under which:
$C \rightarrow e^{- \lambda} C, \bar C \rightarrow e^{+ \lambda} \bar C,
A_\mu \rightarrow A_\mu, B \rightarrow B, {\cal B} \rightarrow {\cal B}$
where $\lambda$ is an infinitesimal global parameter.
According to the Noether's theorem, all the above continuous symmetries
lead to the derivation of conserved charges which turn out to be the
generators for the above transformations. For the generic field $\Psi$,
this statement can be expressed succinctly, in the mathematical form, as
$$
\begin{array}{lcl}
s_{r} \Psi = - i \; [ \Psi, Q_{r} ]_{\pm}, \qquad \;\;\;
r = b, ab, d, ad, w, g, 
\end{array} \eqno(2.5)
$$
where $Q_{r}$ are the conserved charges corresponding to the above symmetries
and brackets $[ \; , \; ]_{\pm}$ stand for the (anti-)commutators for $\Psi$
being (fermionic)bosonic in nature. The local expressions for the charges 
$Q_{r}$ , which are not required for our present discussion,
are given in Refs. [29,30,35] for the case of Feynman gauge where $\xi = 1$
in (2.1) and (2.3).

Together, the above six local, continuous and covariant symmetry 
transformations for the Lagrangian densities (2.1) (and (2.3))
obey the following operator algebra
$$
\begin{array}{lcl}
&& s_{b}^2 = s_{d}^2 = s_{ab}^2 = s_{ad}^2 = 0, \qquad
s_{w} = \{ s_{b}, s_{d} \} = \{  s_{ab}, s_{ad} \}, \nonumber\\
&& s_{d} s_{ad} + s_{ad} s_{d} = 0, \quad
 s_{b} s_{ab} + s_{ab} s_{b} = 0, \quad [s_{w}, s_{r} ] = 0, 
\quad s_{r} = s_{b}, s_{ab}, s_{d},  s_{ad}, s_{g}, \nonumber\\
&&i [s_{g}, s_{b}] = + s_{b}, \quad i [s_{g}, s_{d}] = - s_{d},\quad
i [s_{g}, s_{ab}] = - s_{ab}, \quad i [s_{g}, s_{ad}] 
= +  s_{ad},
\end{array} \eqno(2.6)
$$
which is reminiscent of the algebra obeyed by the de Rham cohomological
operators. Thus, we see that there is a mapping between the cohomological
operators on one hand and symmetry transformations on the other hand. This
mapping is $d \Leftrightarrow (s_{b}, s_{ad}), \delta \Leftrightarrow
(s_{d}, s_{ab}), \Delta \Leftrightarrow \{s_{b}, s_{d} \} = 
\{ s_{ab}, s_{ad} \}$. {\it In physical language, it can be noticed that 
it is the kinetic energy term, gauge-fixing term and  ghost term of the
Lagrangian densities (2.1) and (2.3) that remain invariant under
the (anti-)BRST $s_{(a)b}$, (anti-)co-BRST $s_{(a)d}$ and bosonic symmetry
$s_{w}$ transformations, respectively}. It is straightforward to check
that the Lagrangian densities in (2.1) and (2.3) can be expressed
in terms of transformations in (2.2), (2.4) and (2.5), modulo some 
total derivatives, as
$$
\begin{array}{lcl}
{\cal L}_{b} = \{\; Q_{d}, \frac{1}{2} E C\; \} - \{\; Q_{b}, \frac{1}{2}
(\partial_{\rho} A^\rho) \bar C \;\} \equiv 
\{\; Q_{ab}, \frac{1}{2}
(\partial_{\rho} A^\rho) \;C \;\}
- \{\; Q_{ad}, \frac{1}{2} E \bar C\; \},
\end{array} \eqno(2.7)
$$
$$
\begin{array}{lcl}
{\cal L}_{B} &=& {\cal B} E - \frac{1}{2}\; {\cal B}^2 + X, \;\;\;\qquad\;\;\;\;
{\cal L}_{B} = B (\partial_\rho A^\rho) + \frac{\xi}{2}\; B^2 + Y, 
\nonumber\\
X &=& \tilde s_{b} \tilde s_{ab} \bigl (\; \frac{i}{2} A^2 
- \frac{\xi}{2} \bar C C\; \bigr )
\equiv \tilde s_{b} \bigl (- i \bar C\;
 [\;\partial_\rho A^\rho + \frac{\xi}{2} B\;] \bigr )
\equiv \tilde s_{ab} \bigl (+ i C\; [\;\partial_\rho A^\rho 
+ \frac{\xi}{2} B\;]
 \bigr ), \nonumber\\
Y &=& \tilde s_{d} \tilde s_{ad} \bigl (\; \frac{i}{2} A^2 
- \frac{1}{2} \bar C C\; \bigr )
\equiv \tilde s_{ad} \bigl (- i\; \bar C\; [\; E 
- \frac{1}{2} {\cal B}\; ]\; \bigr )
\equiv \tilde s_{d} \bigl (+ i\; C\; [\; E 
-  \frac{1}{2} {\cal B}\; ] \;\bigr ).
\end{array} \eqno(2.8)
$$
The appearance of the Lagrangian density in (2.7) is like Witten type TFTs
if we assume that the vacuum as well as physical states of the theory 
are invariant under the (anti-)BRST and (anti-)co-BRST symmetries
(i.e. $Q_{(a)b} |vac> = 0, Q_{(a)b} |phys> = 0, Q_{(a)d} |vac> 
= 0, Q_{(a)d} |phys> = 0$). Such a requirement is
satisfied when we choose the physical state (as well as the vacuum) to be
the harmonic state of the Hodge decomposed version of any arbitrary state in
the quantum Hilbert space [30,35]. In contrast to the Witten type
appearance of the Lagrangian density (2.1), the local symmetries of the
theory are that of Schwarz type. Thus, Lagrangian density (2.1) (or (2.3))
describes a new type of TFT [35]. The topological nature of this theory
is confirmed by the following expression for the symmetric energy momentum
tensor $T_{\alpha\beta}^{(s)}$ for the Lagrangian density (2.1)
$$
\begin{array}{lcl}
T_{\alpha\beta}^{(s)} &=& - \frac{1}{2 \xi}\; (\partial_\rho A^\rho)\;
[\; \partial_\alpha A_\beta + \partial_\beta A_\alpha \;] - \frac{1}{2} \; E
\;[\; \varepsilon_{\alpha\rho} \partial_\beta A^\rho 
+ \varepsilon_{\beta\rho} \partial_\alpha A^\rho \;] \nonumber\\
&-& i\; \partial_\alpha \bar C \partial_\beta C 
- i\; \partial_\beta \bar C \partial_\alpha C
- \eta_{\alpha\beta} {\cal L}_{b}  
\end{array} \eqno(2.9)
$$
which turns out to be the sum of (co-)BRST anti-commutators as given below
$$
\begin{array}{lcl}
T_{\alpha\beta}^{(s)} &=& \{ Q_{b}, V_{\alpha\beta}^{(1)} \}
+ \{ Q_{d}, V_{\alpha\beta}^{(2)} \} \equiv
\{ Q_{ab}, \bar V_{\alpha\beta}^{(1)} \}
+ \{ Q_{ad}, \bar V_{\alpha\beta}^{(2)} \} 
\nonumber\\
V_{\alpha\beta}^{(1)} &=& \frac{1}{2} \;\bigl  [ 
\;(\partial_\alpha \bar C) A_\beta +
(\partial_\beta \bar C) A_\alpha + \eta_{\alpha\beta} (\partial_\rho A^\rho)
\bar C \;\bigr ], \nonumber\\
V_{\alpha\beta}^{(2)} &=& \frac{1}{2} \;\bigl  [ 
\;(\partial_\alpha C) \varepsilon_{\beta\rho} A^\rho +
(\partial_\beta C) \varepsilon_{\alpha\rho} A^\rho 
- \eta_{\alpha\beta} E C \;\bigr ],
\end{array} \eqno(2.10)
$$
where $\bar V's$ can be derived from the $V's$ by the substitution 
$C\leftrightarrow \bar C, E \leftrightarrow - E, (\partial_\mu A^\mu)
\leftrightarrow - (\partial_\mu A^\mu)$ under which
$s_{b} \leftrightarrow s_{ab}, s_{d} \leftrightarrow s_{ad}$ in (2.2).
The form of $T_{\alpha\beta}^{(s)}$ in the above equation establishes the fact
that there are no energy excitations ($<phys| T_{00}^{(s)} |phys^\prime> = 0$)
in the physical sector of the theory because the 
{\it hermitian} (co-)BRST charges
annihilate all the physical states. On the 2D compact manifold
\footnote{To be very precise, 2D Minkowski manifold is not a compact manifold. 
For the discussion of the topological invariants, homology cycles, etc. 
and their connections with the curves in the algebraic geometry, one
should consider the Euclidean version of the above manifold which turns 
out to be a 2D closed Riemann surface. Here the metric, unlike the case
of Minkowskian manifold, carries the same 
diagonal signatures and $ \mu, \nu....= 1, 2$. 
For the sake of brevity, however, we shall continue 
with our earlier Minkowskian notations but shall keep in mind this 
important fact and very delicate issue (see, e.g., [47,48]).}, there are
three ($k = 0, 1, 2)$ topological invariants  
(i.e., zero-form, one-form and two-form) which can be generically expressed as
$$
\begin{array}{lcl}
I_{k} = {\displaystyle \oint}_{C_{k}}\; V_{k}, \qquad 
J_{k} = {\displaystyle \oint}_{C_{k}}\; W_{k}, \qquad (k = 0, 1, 2),
\end{array} \eqno(2.11)
$$
where $C_{k}$ are the $k$-dimensional homology cycles in the 2D manifold
and $I_{k}$ and $J_{k}$ are the invariants w.r.t. BRST and co-BRST charges 
respectively. Similar expressions can be obtained (with $ C \leftrightarrow 
\bar C, E \leftrightarrow - E, (\partial_\mu A^\mu) \leftrightarrow
- (\partial_\mu A^\mu)$) as far as 
the nilpotent anti-BRST and anti-co-BRST charges are concerned.
These invariants are connected with one-another by a
certain specific kind of recursion
relations [35]. Thus, if we know the zero-forms (that are 
explicitly BRST- and co-BRST invariants), we can compute the rest 
of the forms by exploiting the recursion relations. For
the Lagrangian densities (2.1) and (2.3), these physical (anti-)BRST- 
and (anti-)co-BRST invariant quantities (zero-forms) are
$$
\begin{array}{lcl}
&& \bar V_{0} = - \frac{1}{\xi}
(\partial_\rho A^\rho) \bar C, \qquad \bar W_{0} = E C, \qquad
\bar {\cal V}_{0} = B \bar C, \qquad \bar {\cal W}_{0} = {\cal B} C,
\nonumber\\
&& V_{0} = - \frac{1}{\xi} (\partial_\rho A^\rho) C, \qquad
W_{0} = E \bar C, \qquad
{\cal V}_{0} = B C, \qquad {\cal W}_{0} =  {\cal B} \bar C, 
\end{array} \eqno(2.12)
$$
where $(\bar V_{0}) V_{0}$ and $(\bar W_{0}) W_{0}$ 
are on-shell ($\Box C = \Box \bar C = 0$) invariant
quantities for the Lagrangian density (2.1) and 
$(\bar {\cal V}_{0}){\cal V}_{0}$
and $(\bar {\cal W}_{0}) {\cal W}_{0}$ are 
off-shell invariant quantities for (2.3).\\

\noindent
{\bf 2.2 On-shell nilpotent (co-)BRST symmetries: 
chiral superfield formalism}\\

\noindent
To provide the geometrical origin and interpretation for the 
on-shell nilpotent (co-)BRST
symmetries and corresponding generators, we resort to the superfield
formulation on the four $(2+2)$-dimensional supermanifold. To this end
in mind, first of all we generalize the generic local field 
$\Psi (x) = (A_\mu (x), C (x), \bar C (x))$ of the Lagrangian density (2.1)
to a chiral ($\partial_{\theta} \tilde A_{M}
(x,\theta,\bar\theta) = 0$) supervector superfield $\tilde A_{M} (x,\bar\theta)
= (B_\mu (x,\bar\theta), \Phi (x,\bar \theta), \bar \Phi (x,\bar\theta))$
with the following super expansions along the Grassmannian 
$\bar\theta$-direction of the $(2+2)$-dimensional supermanifold
$$
\begin{array}{lcl}
B_{\mu} (x,  \bar \theta) &=& A_{\mu} (x) 
+\; \bar \theta\; R_{\mu} (x), \nonumber\\ 
\Phi (x,  \bar \theta) &=& C (x) 
- i\; \bar \theta \;{\cal B} (x), \nonumber\\
\bar \Phi (x,  \bar \theta) &=& \bar C (x) 
+ i\; \bar \theta \;B (x). 
\end{array} \eqno(2.13)
$$
There are a few salient points which we summon here: (i) it is
obvious that in the limit $\bar \theta \rightarrow 0$, we get back the
generic field $\Psi (x)$ of the Lagrangian density (2.1). (ii) In general,
a superfield on the four $(2+2)$-dimensional supermanifold is parametrized by
the superspace variables $Z^M = (x^\mu, \theta,\bar\theta)$ where
$x^\mu (\mu = 0, 1)$ are the two even spacetime variables and $\theta, 
\bar\theta$ are the odd variables (with $\theta^2 = \bar\theta^2 = 0,
\theta\bar\theta = - \bar\theta \theta$). However, for our present discussions,
we have chosen $Z^M = (x^\mu, \bar\theta)$. (iii) The minus sign in the
super expansion of $\Phi (x,\bar\theta)$ has been taken just for the
algebraic convenience. (iv) The total number of degrees of freedom for
the odd fields $(R_\mu, C, \bar C)$ and even fields $(A_\mu, B, {\cal B})$
match in the above expansion for the sake of consistency with the basic
tenets of supersymmetry. (v) The auxiliary fields $R_\mu, B, {\cal B}$ will be 
fixed in terms of the basic fields after the application of the (dual)
horizontality conditions. Some of them can also be fixed by resorting to the 
equations of motion for the Lagrangian density (2.3). (vi) All the component
fields, on the r.h.s. of the above expansion, are functions of the spacetime
even variable $x^\mu$ alone. (vii) The super expansions in (2.13) are the
chiral limit ($\theta \rightarrow 0$) of the most general expansion of the
superfields on the $(2+2)$-dimensional supermanifold (see, e.g., [38])
$$
\begin{array}{lcl}
B_{\mu} (x, \theta, \bar \theta) &=& A_{\mu} (x) 
+ \theta\; \bar R_{\mu} (x) + \bar \theta\; R_{\mu} (x) 
+ i \;\theta \;\bar \theta S_{\mu} (x), \nonumber\\
\Phi (x, \theta, \bar \theta) &=& C (x) 
- i\; \theta  B (x)
-i \;\bar \theta\; {\cal B} (x) 
+ i\; \theta\; \bar \theta \;s (x), \nonumber\\
\bar \Phi (x, \theta, \bar \theta) &=& \bar C (x) 
+ i \;\theta\; {\cal B} (x) + i\; \bar \theta \;B (x) 
+ i \;\theta \;\bar \theta \;\bar s (x),
\end{array} \eqno(2.14)
$$
where signs in the above expansion have been chosen for the algebraic
convenience and auxiliary fields $B$ and ${\cal B}$ are the ones present in
the Lagrangian density (2.3). Here too, it can be seen that the fermionic
$(R_\mu, \bar R_\mu, C , \bar C, s, \bar s)$ degrees of freedom do match with
that of the bosonic $(A_\mu, S_\mu, + B, - B, +{\cal B}, - {\cal B})$ degrees.

Now we exploit the horizontality condition 
($\tilde F = \tilde d \tilde A = d A = F$) w.r.t. (super) exterior 
derivatives $(\tilde d) d$ in the construction of the (super) curvature
two-form. Physically, the requirement of the horizontality condition
implies an imposition that the curvature two-form in the ordinary
spacetime manifold remains unchanged and it is restricted not to get any
contribution from the {\it superspace} variables.
The explicit expressions for the l.h.s. and r.h.s. in 
$(\tilde d \tilde A = d A$) are
$$
\begin{array}{lcl}
\tilde d \tilde A &=& (dx^\mu \wedge dx^\nu) (\partial_\mu B_\nu)
+ (dx^\mu \wedge d \bar\theta) ( \partial_\mu \Phi - \partial_{\bar \theta}
B_\mu ) - (d \bar\theta \wedge d \bar \theta) (\partial_{\bar\theta} \Phi),
\nonumber\\
d A &=& (dx^\mu \wedge dx^\nu)\; (\partial_\mu A_\nu) 
\equiv \frac{1}{2} (dx^\mu \wedge dx^\nu) (\partial_\mu A_\nu 
- \partial_\nu A_\mu),
\end{array} \eqno(2.15)
$$
where the super exterior derivative $\tilde d$ (defined in terms of the chiral
superspace coordinates) and super connection one-form $\tilde A$
(defined in terms of the chiral superfields) are
$$
\begin{array}{lcl}
\tilde d &=& d Z^M \;\partial_{M} \equiv
d x^\mu \;\partial_\mu + d \bar \theta\;\partial_{\bar\theta},
\nonumber\\
\tilde  A &=& d Z^M\; \tilde A_{M} \equiv
dx^\mu \;B_\mu (x,\bar\theta) 
+ d \bar \theta\; \Phi (x,\bar\theta).
\end{array} \eqno(2.16)
$$
The above equations are the chiral limits ($\theta \rightarrow 0,
d \theta \rightarrow 0$) of the following most general definitions
for the super exterior derivative and super one-form connection 
(see, e.g., [38])
$$
\begin{array}{lcl}
\tilde d &=& d Z^M \;\partial_{M} \equiv
d x^\mu \;\partial_\mu \;+\; d \bar \theta\;\partial_{\bar\theta} \;+\; 
d \theta \;\partial_{\theta},
\nonumber\\
\tilde  A &=& d Z^M\; \tilde A_{M} \equiv
dx^\mu \;B_\mu (x,\theta,\bar\theta) 
+ d \bar \theta\; \Phi (x,\theta,\bar\theta) 
+ d \theta\; \bar \Phi (x,\theta,\bar\theta),
\end{array} \eqno(2.17)
$$
on the $(2+2)$-dimensional compact supermanifold. It is straightforward to 
check that the horizontality restriction $\tilde d \tilde A = d A$ leads to
the following relationships
$$
\begin{array}{lcl}
\partial_{\bar\theta} \Phi (x,\bar\theta)
= 0 \rightarrow {\cal B} (x) = 0, \;\qquad\;\;
\partial_{\bar\theta} B_\mu (x,\bar\theta) = \partial_\mu \Phi (x,\bar\theta)
\rightarrow 
R_\mu (x) = \partial_\mu C (x),
\end{array} \eqno(2.18)
$$
and the restriction $\partial_\mu B_\nu - \partial_\nu B_\mu =
\partial_\mu A_\nu - \partial_\nu A_\mu $ implies
$\partial_\mu R_\nu - \partial_\nu R_\mu = 0$ which is readily satisfied by
$R_\mu = \partial_\mu C$. It is obvious that the condition $\tilde d \tilde A
= d A$ does not fix the auxiliary field $B(x)$ in terms of the basic fields
of the Lagrangian density (2.1). However, the equation of motion for the
Lagrangian density (2.3) comes to our rescue as: $B (x) = - \frac{1}{\xi}
(\partial_\mu A^\mu) (x)$. With these substitutions for the auxiliary fields, 
the super expansion (2.13) becomes:
$$
\begin{array}{lcl}
B_{\mu} (x,  \bar \theta) &=& A_{\mu} (x) 
+\; \bar \theta\; \partial_{\mu} C (x)
\equiv A_\mu (x) + \bar\theta\; (s_{b} A_\mu (x)), \nonumber\\ 
\Phi (x,  \bar \theta) &=& C (x) 
- i\; \bar \theta \;({\cal B} (x) = 0)
\equiv C (x) + \bar\theta\; (s_{b} C (x) = 0), \nonumber\\
\bar \Phi (x,  \bar \theta) &=& \bar C (x) 
- \frac{i}{\xi} \; \bar \theta \;(\partial_\mu A^\mu) (x)
\equiv \bar C (x) + \bar\theta\; (s_{b} \bar C(x)). 
\end{array} \eqno(2.19)
$$
In fact, now the on-shell nilpotent BRST symmetry transformations in (2.2)
can be concisely written in terms of the above superfields expansions as
$$
\begin{array}{lcl}
s_{b} B_\mu (x,\bar\theta) = \partial_\mu \Phi (x,\bar\theta),\quad 
s_{b} \Phi (x,\bar\theta) = 0,\quad 
s_{b} \bar \Phi (x,\bar\theta) = - \frac{i}{\xi} 
(\partial_\mu B^\mu) (x,\bar\theta).
\end{array} \eqno(2.20)
$$
One can readily check that the first transformation in the above equation 
leads to $s_{b} A_\mu = \partial_\mu C, s_{b} C = 0$; the second 
transformation produces
$s_{b} C = 0$ and the third one generates $s_{b} \bar C = -\frac{i}{\xi}
(\partial_\mu A^\mu), s_{b} (\partial_\mu A^\mu) = \Box C$ in terms of the
basic fields of Lagrangian density (2.1).
It is interesting to check, vis-a-vis equation (2.5),  that
$$
\begin{array}{lcl}
{\displaystyle \frac{\partial}{\partial \bar\theta}}\; 
\tilde A_{M} (x,\bar\theta)
= - i \;\bigl [ \Psi (x), Q_{b} \bigr ]_{\pm} \equiv s_{b} \Psi (x), \quad
\tilde A_{M} = (\Phi, \bar \Phi, B_\mu), \quad \Psi = (C , \bar C, A_\mu),
\end{array} \eqno(2.21)
$$
where the brackets $[\;,\;]_{\pm}$ stand for the (anti-)commutator when the
generic field $\Psi$ and superfield $\tilde A_M$ are (fermionic)bosonic
in nature. {\it Thus, conserved and nilpotent BRST charge $Q_{b}$ 
geometrically turns out to be 
the translation generator $\partial / \partial \bar \theta$ 
for the superfields $\tilde A_M$ along the $\bar\theta$-direction 
of the supermanifold. The process of this translation generates
the on-shell nilpotent BRST symmetry transformations on $\Psi$
which correspond to (2.2).} In addition, the nilpotency of $s_{b}^2 = 0$
(and $Q_{b}^2 = 0$) is intimately connected with the property of the
square of the translational generator 
(i.e. $(\partial/\partial\bar\theta)^2 = 0$).

Now we illustrate the derivation of the on-shell nilpotent dual(co-)BRST
symmetry transformations of (2.2) by exploiting the analogue of the 
horizontality condition 
\footnote{ We christen this condition as the dual horizontality condition
because $\tilde d (d)$ and $\tilde \delta (\delta)$ are dual 
($\tilde \delta = - \star \tilde d \star, \delta = - * d *$) 
to each-other. The restriction $\tilde \delta \tilde A = \delta A$ amounts
to setting equal to zero all the Grassmannian parts of the superscalar
$\tilde \delta \tilde A$. On an ordinary even dimensional manifold, the 
operation $\delta A = - * d * A $ always yields the (zero-form) 
covariant gauge-fixing
term (i.e. $\delta A = \partial_\mu A^\mu$) for the one-form ($A (x) 
= dx^\mu A_\mu (x))$ Abelian $U(1)$ gauge theory in any arbitrary
spacetime dimension.} w.r.t. (super) 
co-exterior derivatives 
$(\tilde\delta) \delta$ by requiring $\tilde \delta \tilde A = \delta A$ 
with the following expressions 
$$
\begin{array}{lcl}
\tilde \delta \tilde A &=& \partial_\mu B^\mu + s^{\bar\theta\bar\theta}
(\partial_{\bar\theta} \bar \Phi) - \varepsilon^{\mu\bar\theta}
(\partial_\mu \bar\Phi + \varepsilon_{\mu\nu} \partial_{\bar\theta} B^\nu),
\nonumber\\
\delta A &=& \partial_\mu A^\mu, \qquad A (x) = dx^\mu A_\mu (x), \qquad
\tilde A = dZ^M \tilde A_{M},
\end{array} \eqno(2.22)
$$
where $\delta = - * d *$,  $\tilde \delta = - \star \tilde d \star$ and
$\star$ corresponds to the 
Hodge duality operation on the compact $(2+2)$-dimensional
supermanifold. Physically, the dual horizontality condition implies
a restriction on the zero-form gauge-fixing term ($\partial_\mu A^\mu$),
defined on the ordinary spacetime manifold, to remain unchanged. In other
words, the {\it superspace} contribution to the gauge-fixing term is
required to be zero. The operation of $\star$ on the superspace  differentials
$ d Z^M$ and their wedge products $(d Z^M \wedge d Z^N)$, in the most
general form, are
$$
\begin{array}{lcl}
\star\; (dx^\mu) &=& \varepsilon^{\mu\nu}\; (d x_\nu), \qquad \;
\star\; (d\theta) = (d \bar \theta), \qquad \;
\star\; (d \bar \theta) = (d \theta), \nonumber\\
\star\; (d x^\mu \wedge d x^\nu) &=& \varepsilon^{\mu\nu}, \qquad 
\star\; (dx^\mu \wedge d \theta) = \varepsilon^{\mu\theta}, \qquad
\star\; (dx^\mu \wedge d \bar \theta) = \varepsilon^{\mu\bar\theta},
\nonumber\\
\star \; (d \theta \wedge d \theta) &=& s^{\theta\theta}, \qquad
\star \; (d \theta \wedge d \bar \theta) = s^{\theta \bar \theta}, \qquad
\star \; (d \bar \theta \wedge d \bar \theta) = s^{\bar \theta \bar \theta},
\end{array} \eqno(2.23)
$$
where $\varepsilon^{\mu\theta} = - \varepsilon^{\theta\mu}, \varepsilon^{\mu
\bar\theta} = - \varepsilon^{\bar\theta \mu},  
s^{\theta \bar \theta} = s^{\bar\theta\theta}$ etc. It will be noted that, in
the derivation of $\tilde \delta \tilde A = - \star \tilde d \star\; 
\tilde A$, we have used the following expansion
$$
\begin{array}{lcl}
\star \tilde A = \varepsilon^{\mu\nu} (dx_\nu) B_\mu (x,\bar\theta)
+ d \bar\theta \; \bar \Phi (x,\bar\theta),
\end{array} \eqno(2.24)
$$
which emerges from the chiral limit of the $\star \tilde A$ derived from the
most general definition of $\tilde A = d Z^M \tilde A_{M}$ in (2.17) and 
application of the $\star$ operation (2.23) on it. The dual
horizontality condition w.r.t. $(\tilde \delta)\delta$
(i.e. $\tilde \delta \tilde A = \delta A$) leads to
$$
\begin{array}{lcl}
\partial_{\bar\theta} \bar \Phi = 0 \rightarrow B (x) = 0, \qquad
\partial_\mu \bar \Phi = - \varepsilon_{\mu\nu} \partial_{\bar\theta} B^\nu
\rightarrow R_\mu = - \varepsilon_{\mu\nu} \partial^\nu \bar C,
\end{array} \eqno(2.25)
$$
and $\partial_\mu B^\mu = \partial_\mu A^\mu$ implies $\partial_\mu R^\mu = 0$
that is satisfied automatically by $R_\mu = - \varepsilon_{\mu\nu} 
\partial^\nu \bar C$. It is obvious that the auxiliary field ${\cal B} (x)$
is not fixed in terms of the basic fields of the Lagrangian density (2.1)
by the dual horizontality condition
(i.e. $\tilde \delta \tilde A = \delta A$). However, the equation of motion
for the Lagrangian density (2.3) helps us to get ${\cal B} = E$. Thus, the
chiral super expansion (2.13), on the chiral supermanifold,  becomes
$$
\begin{array}{lcl}
B_{\mu} (x,  \bar \theta) &=& A_{\mu} (x) 
-\; \bar \theta\; \varepsilon_{\mu\nu} \partial^{\nu} \bar C (x)
\equiv A_\mu (x) + \bar\theta\; (s_{d} A_\mu (x)), \nonumber\\ 
\Phi (x,  \bar \theta) &=& C (x) 
- i\; \bar \theta \; E (x)
\equiv C (x) + \bar\theta\; (s_{d} C (x)), \nonumber\\
\bar \Phi (x,  \bar \theta) &=& \bar C (x) 
+ i \; \bar \theta \;(B (x) = 0)
\equiv \bar C (x) + \bar\theta\; (s_{d} \bar C(x) = 0). 
\end{array} \eqno(2.26)
$$
In terms of the above chiral superfield expansion, the dual(co-)BRST
symmetry transformations of (2.2) can be concisely expressed as
$$
\begin{array}{lcl}
s_{d} B_\mu (x,\bar\theta) = - \varepsilon_{\mu\nu}
\partial^\nu \bar \Phi (x,\bar\theta),\quad 
s_{d} \bar \Phi (x,\bar\theta) = 0,\quad 
s_{b} \Phi (x,\bar\theta) = + i \varepsilon^{\mu\nu} \partial_\mu B_\nu.
\end{array} \eqno(2.27)
$$
It is now evident that
$$
\begin{array}{lcl}
{\displaystyle \frac{\partial}{\partial \bar\theta}}\; 
\tilde A_{M} (x,\bar\theta)
= - i \;\bigl [ \Psi (x), Q_{d} \bigr ]_{\pm} \equiv s_{d} \Psi (x), \quad
\tilde A_{M} = (\Phi, \bar \Phi, B_\mu), \quad \Psi = (C , \bar C, A_\mu),
\end{array} \eqno(2.28)
$$
where the brackets have the same meaning as discussed earlier.
This equation shows that {\it geometrically} the on-shell nilpotent co-BRST
charge $Q_{d}$ is the generator of translation $\partial/ \partial \bar\theta$
for the chiral superfield $\tilde A_M$ along the Grassmannian direction
$\bar\theta$ of the $(2+2)$-dimensional supermanifold. Furthermore, the 
on-shell nilpotency conditions $s_{d}^2 = 0, Q_{d}^2 = 0$ are connected
with the property of the square of the translational generator 
$(\partial/\partial \bar\theta)^2 = 0$. The process of
the translation of $\tilde A_M (x,\bar\theta)
= (B_\mu, \Phi, \bar \Phi) (x,\bar\theta) $ along the $\bar\theta$-direction 
produces the co-BRST transformation $s_{d} \Psi$ for the
generic local field $\Psi = (A_\mu, C, \bar C)$
(i.e. $s_{d} \leftrightarrow \partial/\partial\bar\theta$). 
{\it There is a clear 
distinction, however, between $Q_{b}$ and $Q_{d}$ as far as translation of
the fermionic superfields (or transformations on (anti-)ghost fields 
$(\bar C)C$) along $\bar\theta$-direction is concerned. For instance,
the translation generated by $Q_{b}$ along $\bar\theta$-direction
results in the transformation for the anti-ghost field
$\bar C$ but analogous translation by $Q_{d}$ leads to the transformation 
for the ghost field $C$}. In more sophisticated language, the horizontality
condition entails upon the chiral superfield $\bar \Phi$ to remain chiral
but the chiral superfield $\Phi$ becomes a local spacetime field 
(i.e., $\Phi (x,\bar\theta) = C (x)$). On the contrary, the dual horizontality
condition entails upon the chiral
superfield $\Phi$ to retain its chirality but 
the chiral superfield $\bar\Phi$ becomes 
an ordinary local field (i.e., $\bar\Phi (x,\bar\theta) = \bar C (x)$).\\

\noindent
{\bf 2.3 Anti-BRST and anti-co-BRST symmetries: anti-chiral superfields}\\

\noindent
To derive the on-shell nilpotent anti-BRST and anti-co-BRST symmetry
transformations of (2.2), we resort to the anti-chiral superfields
$\tilde A_M (x,\theta) = (B_\mu, \Phi, \bar \Phi) (x,\theta)$ which have
the following super expansions along the $\theta$-direction of the
anti-chiral supermanifold
$$
\begin{array}{lcl}
B_{\mu} (x,  \theta) &=& A_{\mu} (x) 
+\; \theta\; \bar R_{\mu} (x), \nonumber\\ 
\Phi (x,  \bar \theta) &=& C (x) 
- i\; \theta \;B (x), \nonumber\\
\bar \Phi (x,  \theta) &=& \bar C (x) 
+ i\; \theta \;{\cal B} (x). 
\end{array} \eqno(2.29)
$$
These are, in fact, the anti-chiral limit ($\bar\theta \rightarrow 0$) 
of the general super expansion (2.14) on the
$(2+2)$-dimensional supermanifold. The super exterior derivative $\tilde d$
and super connection one-form $\tilde A$, for our present discussion, are
$$
\begin{array}{lcl}
\tilde d &=& d Z^M \;\partial_{M} \equiv
d x^\mu \;\partial_\mu + d  \theta\;\partial_{\theta},
\nonumber\\
\tilde  A &=& d Z^M\; \tilde A_{M} \equiv
dx^\mu \;B_\mu (x,\theta) 
+ d  \theta\; \bar \Phi (x,\theta),
\end{array} \eqno(2.30)
$$
which are the anti-chiral limit ($\theta \rightarrow 0, 
d \theta \rightarrow 0$) of the corresponding general expressions defined 
in (2.17). Now the imposition of the horizontality condition ($\tilde d 
\tilde A = d A)$ implies the restriction that the curvature two-form
$F = d A$, defined on the ordinary spacetime manifold, remains unchanged.
In other words, the superspace contributions to the curvature
two-form are set equal to zero. With the above definitions (2.30)
on the anti-chiral supermanifold, the following inputs
$$
\begin{array}{lcl}
\tilde d \tilde A &=& (dx^\mu \wedge dx^\nu) (\partial_\mu B_\nu)
+ (dx^\mu \wedge d \theta) ( \partial_\mu \bar \Phi - \partial_{ \theta}
B_\mu ) - (d \theta \wedge d \theta) (\partial_{\theta} \bar \Phi),
\nonumber\\
d A &=& (dx^\mu \wedge dx^\nu)\; (\partial_\mu A_\nu) \equiv
\frac{1}{2}\; (dx^\mu \wedge dx^\nu)\; (\partial_\mu A_\nu - \partial_\nu
A_\mu), 
\end{array} \eqno(2.31)
$$
lead to the following relationships due to $d A = \tilde d \tilde A$
$$
\begin{array}{lcl}
\partial_{\theta} \bar \Phi (x,\theta)
= 0 \rightarrow {\cal B} (x) = 0, \qquad
\partial_\mu \bar \Phi (x,\theta)
= \partial_\theta B_\mu (x,\theta) \rightarrow \bar R_\mu (x)
= \partial_\mu \bar C (x), 
\end{array} \eqno(2.32)
$$
and $\partial_\mu B_\nu - \partial_\nu B_\mu =
\partial_\mu A_\nu - \partial_\nu A_\mu $  which implies
$\partial_\mu \bar R_\nu - \partial_\nu \bar R_\mu = 0$. The latter requirement
is automatically satisfied by $\bar R_\mu = \partial_\mu \bar C$. It is clear
that the above horizontality restriction does not fix $B(x)$ in terms of the
basic fields of the Lagrangian density (2.1). However, the equation of
motion $ B = - \frac{1}{\xi}\;
(\partial_\mu A^\mu)$ for the Lagrangian density (2.3) comes
to our help. With these insertions, the super expansion (2.29) becomes
$$
\begin{array}{lcl}
B_{\mu} (x,  \theta) &=& A_{\mu} (x) 
+\; \theta\; \partial_{\mu} \bar C (x)
\equiv A_\mu (x) + \theta\; (s_{ab} A_\mu (x)), \nonumber\\ 
\Phi (x,  \bar \theta) &=& C (x) 
+ \frac{i}{\xi}\; \theta \; (\partial_\mu A^\mu) (x)
\equiv C (x) + \theta\; (s_{ab} C (x)), \nonumber\\
\bar \Phi (x,  \theta) &=& \bar C (x) 
+ i\; \theta \;({\cal B} (x) = 0)
\equiv \bar C (x) + \theta \; (s_{ab} \bar C (x) = 0). 
\end{array} \eqno(2.33)
$$
In terms of the above anti-chiral superfield expansions, the anti-BRST
symmetry transformations of (2.2) can be concisely expressed as
$$
\begin{array}{lcl}
s_{ab} B_\mu (x,\theta) = 
\partial_\mu \bar  \Phi (x,\theta),\quad 
s_{ab} \bar \Phi (x,\theta) = 0,\quad 
s_{ab} \Phi (x,\theta) = + \frac{i}{\xi} (\partial_\mu B^\mu) (x,\theta).
\end{array} \eqno(2.34)
$$
It is now straightforward to check that
$$
\begin{array}{lcl}
{\displaystyle \frac{\partial}{\partial \theta}}\; \tilde A_{M} (x,\theta)
= - i \;\bigl [ \Psi (x), Q_{ab} \bigr ]_{\pm} \equiv s_{ab} \Psi (x), \;
\tilde A_{M} = (\Phi, \bar \Phi, B_\mu), \; \Psi = (C , \bar C, A_\mu),
\end{array} \eqno(2.35)
$$
where the above brackets have the same interpretation as discussed earlier.
This equation shows that {\it geometrically} the on-shell nilpotent anti-BRST
charge $Q_{ab}$ is the generator of translation 
$\frac{\partial} {\partial \theta}$
for the anti-chiral superfield $\tilde A_{M} (x,\theta) = 
(B_\mu, \Phi, \bar\Phi) (x,\theta)$
along the $\theta$-direction of the supermanifold. In fact, this process of
translation produces the anti-BRST symmetry transformations 
(i.e. $s_{ab} \Psi$) for the local fields $\Psi$ that are listed in equation 
(2.2). Thus, there is a mapping 
$s_{ab} \leftrightarrow \frac{\partial}{\partial\theta}$ 
between the above two key operators and the nilpotency
$s_{ab}^2 = 0 (Q_{ab}^2 = 0)$ is encoded in the square of the translation 
generator $(\partial/\partial\theta)^2 = 0$.

Now we shall dwell a bit on the derivation of the on-shell nilpotent
anti-co-BRST symmetry in the framework of anti-chiral superfield formulation.
We exploit here the (super) co-exterior derivatives $(\tilde \delta)\delta$
and (super) connection one-forms $(\tilde A) A$ for the dual
horizontality condition
$$
\begin{array}{lcl}
\tilde \delta \tilde A &=& \delta A, \quad \delta = - * d *, \quad
A (x) = dx^\mu A_\mu (x), \quad \delta A = \partial_\mu A^\mu, \nonumber\\
\tilde \delta \tilde A &=& \partial_\mu B^\mu - \varepsilon^{\mu\theta}
(\partial_\mu \Phi + \varepsilon_{\mu\nu} \partial_\theta B^\nu)
+ s^{\theta\theta} \partial_{\theta} \Phi,
\end{array} \eqno(2.36)
$$
where in the computation of the $\tilde \delta \tilde A$, we have used
$\tilde \delta = - \star \tilde d \star$ and the expression for
$$
\begin{array}{lcl}
\star \tilde A = \varepsilon^{\mu\nu}\; (dx_\nu) \;B_\mu (x,\theta)
+ d \theta\;\Phi (x,\theta).
\end{array} \eqno(2.37)
$$
The above equation emerges as the anti-chiral limit 
of the most general form of $\star \tilde A$
$$
\begin{array}{lcl}
\star \tilde A = \varepsilon^{\mu\nu}\; (dx_\nu) \;B_\mu (x,\theta, \bar\theta)
+ d \theta\;\Phi (x,\theta,\bar\theta) + d \bar\theta\; \bar \Phi (x,\theta,
\bar\theta).
\end{array} \eqno(2.38)
$$
The restriction $\tilde \delta \tilde A = \delta A$ 
(which physically implies an imposition that the zero-form gauge-fixing term
$\delta A = \partial_\mu A^\mu$, defined on the ordinary 
spacetime manifold, remains unchanged) leads to the following
relationships
$$
\begin{array}{lcl}
\partial_{\theta}  \Phi = 0 \rightarrow B (x) = 0, \qquad
\partial_\mu \Phi = - \varepsilon_{\mu\nu} \partial_\theta B^\nu 
\rightarrow \bar R_\mu = - \varepsilon_{\mu\nu} \partial^\nu C, 
\end{array} \eqno(2.39)
$$
and $\partial_\mu B^\mu = \partial_\mu A^\mu$ implies $\partial_\mu \bar R^\mu
= 0$ which is readily satisfied by $\bar R_\mu = - \varepsilon_{\mu\nu}
\partial^\nu C$. The dual horizontality condition
$\tilde \delta \tilde A = \delta A$ does not fix the field ${\cal B} (x)$
in terms of the basic fields. The equation of motion ${\cal B} = E$ for
the Lagrangian density (2.3), however, comes to our rescue. The super expansion
with the above insertions turns out to be
$$
\begin{array}{lcl}
B_{\mu} (x,  \theta) &=& A_{\mu} (x) 
-\; \theta\; \varepsilon_{\mu\nu} \partial^{\nu} C (x)
\equiv A_\mu (x) + \theta\; (s_{ad} A_\mu (x)), \nonumber\\ 
\Phi (x,  \theta) &=& C (x) 
- i \theta\; (B (x)= 0)
\equiv C (x) + \theta\; (s_{ad} C (x) = 0), \nonumber\\
\bar \Phi (x,  \theta) &=& \bar C (x) 
+ i\; \theta \;E (x)
\equiv \bar C (x) + \theta \; (s_{ad} \bar C (x)). 
\end{array} \eqno(2.40)
$$
With the above expansions as the backdrop, we can express now the anti-co-BRST
transformations of (2.2) in terms of the anti-chiral superfields as
$$
\begin{array}{lcl}
s_{ad} B_\mu (x,\theta) = - \varepsilon_{\mu\nu}
\partial^\nu   \Phi (x,\theta),\quad 
s_{ad}  \Phi (x,\theta) = 0,\quad 
s_{ad} \bar \Phi (x,\theta) = - i\; 
\varepsilon^{\mu\nu} \partial_\mu B_\nu (x,\theta).
\end{array} \eqno(2.41)
$$
The geometrical interpretation for the co-BRST charge $Q_{ad}$ is encoded in 
$$
\begin{array}{lcl}
{\displaystyle \frac{\partial}{\partial \theta}}\; \tilde A_{M} (x,\theta)
= - i \;\bigl [ \Psi (x), Q_{ad} \bigr ]_{\pm} \equiv s_{ad} \Psi (x), \;
\tilde A_{M} = (\Phi, \bar \Phi, B_\mu), \; \Psi = (C , \bar C, A_\mu),
\end{array} \eqno(2.42)
$$
where the brackets $[\;,\;]_{\pm}$ have the same interpretation as explained
earlier. It is obvious to note that $Q_{ad}$ turns out to be the 
translation generator $\frac{\partial}{\partial\theta}$ 
for the anti-chiral superfields
$\tilde A_M (x,\theta) = (B_\mu, \Phi, \bar \Phi) (x,\theta)$ along the
$\theta$-direction of the supermanifold. The action of the
on-shell nilpotent transformation operator $s_{ad}$ on the local fields 
$\Psi$ and the operation of $\frac{\partial}{\partial\theta}$ 
on the anti-chiral superfields
$\tilde A_M (x,\theta)$ are inter-related and there exists a mapping
$s_{ad} \leftrightarrow \frac{\partial}{\partial\theta}$
with $s_{ad}^2= 0$ connected to $(\partial/\partial\theta)^2 = 0$. 
{\it Even though both the charges
$Q_{ad}, Q_{ab}$ have the similar kind of mapping with the translation 
generator, there is a clear distinction between them. Whereas the former
generates a transformation for the ghost field $C$ through the translation
of the superfield $\Phi$, the latter generates
the corresponding transformation on the anti-ghost field $\bar C$ through
the translation of $\bar \Phi$ superfield. The direction of
translation is common for both of them (i.e. the $\theta$-direction of the 
supermanifold)}.

It should be emphasized that the on-shell nilpotent (anti-)BRST-
and (anti-)co-BRST symmetries can be derived {\it together} if we
merge systematically the (anti-)chiral superfields and have the super
expansion as given below [37]
$$
\begin{array}{lcl}
B_{\mu} (x, \theta, \bar \theta) &=& A_{\mu} (x) 
+ \theta\; \bar R_{\mu} (x) + \bar \theta\; R_{\mu} (x) 
+ i \;\theta \;\bar \theta S_{\mu} (x), \nonumber\\
\Phi (x, \theta, \bar \theta) &=& C (x) 
+ \frac{i}{\xi}\; \theta  (\partial_\mu A^\mu) (x)
-i \;\bar \theta\; E (x) 
+ i\; \theta\; \bar \theta \;s (x), \nonumber\\
\bar \Phi (x, \theta, \bar \theta) &=& \bar C (x) 
+ i \;\theta\; E (x) - \frac{ i}{\xi} \; \bar \theta \;(\partial_\mu A^\mu) (x) 
+ i \;\theta \;\bar \theta \;\bar s (x).
\end{array} \eqno(2.43)
$$
In our work [37], these super expansions together with the definitions
in (2.17) and $\star$ operation in (2.23) have been exploited in the 
horizontality condition ($\tilde F = \tilde d \tilde A = d A = F$) which leads 
to the derivation of the auxiliary fields in terms of the basic fields of
the Lagrangian density (2.1). Ultimately, the above super expansions are
expressed in terms of on-shell nilpotent (anti-)BRST symmetries (2.2) as
$$
\begin{array}{lcl}
B_{\mu} (x, \theta, \bar \theta) &=& A_{\mu} (x) 
+ \theta\; (s_{ab} A_\mu (x)) + \bar \theta\; (s_{b} A_\mu (x)) 
+  \;\theta \;\bar \theta (s_{b} s_{ab} A_\mu (x)), \nonumber\\
\Phi (x, \theta, \bar \theta) &=& C (x) 
+ \; \theta  (s_{ab} C (x))
+ \;\bar \theta\; (s_{b} C (x)) 
+ \; \theta\; \bar \theta \;(s_{b} s_{ab} C (x)), \nonumber\\
\bar \Phi (x, \theta, \bar \theta) &=& \bar C (x) 
+ \;\theta\; (s_{ab} \bar C (x)) + \; \bar \theta \;(s_{b} \bar C (x)) 
+  \;\theta \;\bar \theta \;(s_{b} s_{ab} \bar C(x)).
\end{array} \eqno(2.44)
$$
In a similar fashion, the dual horizontality condition
w.r.t. (super) co-exterior derivatives (i.e. $\tilde \delta \tilde A
= \delta A$) have been performed in [37] which finally enforces
(2.43) to be expressed in terms of
the on-shell nilpotent (anti-)co-BRST symmetry transformations (2.2), as
$$
\begin{array}{lcl}
B_{\mu} (x, \theta, \bar \theta) &=& A_{\mu} (x) 
+ \theta\; (s_{ad} A_\mu (x)) + \bar \theta\; (s_{d} A_\mu (x)) 
+  \;\theta \;\bar \theta (s_{d} s_{ad} A_\mu (x)), \nonumber\\
\Phi (x, \theta, \bar \theta) &=& C (x) 
+ \; \theta  (s_{ad} C (x))
+ \;\bar \theta\; (s_{d} C (x)) 
+ \; \theta\; \bar \theta \;(s_{d} s_{ad} C (x)), \nonumber\\
\bar \Phi (x, \theta, \bar \theta) &=& \bar C (x) 
+ \;\theta\; (s_{ad} \bar C (x)) + \; \bar \theta \;(s_{d} \bar C (x)) 
+  \;\theta \;\bar \theta \;(s_{d} s_{ad} \bar C(x)).
\end{array} \eqno(2.45)
$$
We would like to lay stress on the fact that {\it it is only for the free
2D Abelian gauge theory that (anti-)chiral superfields are merged together
systematically to produce the on-shell nilpotent (anti-)BRST and 
(anti-)co-BRST symmetries together. The same is not true for the non-Abelian
gauge theory as we shall see later.}\\

\noindent
{\bf 2.4 Off-shell nilpotent symmetries: superfield approach}\\

\noindent
We shall discuss here the bare essentials of our earlier work
[38] where the off-shell
nilpotent (anti-)BRST and (anti-)co-BRST symmetries have been obtained
by exploiting the most general super expansion of the multiplets of supervector 
superfield $\tilde A_M (x,\theta,\bar\theta) = (B_\mu, \Phi, \bar \Phi)
(x,\theta,\bar\theta)$ as given below
$$
\begin{array}{lcl}
B_{\mu} (x, \theta, \bar \theta) &=& A_{\mu} (x) 
+ \theta\; \bar R_{\mu} (x) + \bar \theta\; R_{\mu} (x) 
+ i \;\theta \;\bar \theta \;S_{\mu} (x), \nonumber\\
\Phi (x, \theta, \bar \theta) &=& C (x) 
+ i\; \theta  \bar B (x)
- i \;\bar \theta\; {\cal B} (x) 
+ i\; \theta\; \bar \theta \;s (x), \nonumber\\
\bar \Phi (x, \theta, \bar \theta) &=& \bar C (x) 
- i \;\theta\; \bar {\cal B} (x) 
+ i \; \bar \theta \;B (x) 
+ i \;\theta \;\bar \theta \;\bar s (x),
\end{array} \eqno(2.46)
$$
where signs are chosen for the algebraic convenience only.
The application of the horizontality condition ($\tilde d \tilde A = d A$)
w.r.t. the (super) exterior
derivatives $(\tilde d) d$ and (super) connection one-forms
$(\tilde A) A$ (cf. (2.17)) leads to the following relationships
$$
\begin{array}{lcl}
&&s(x) = \bar s(x) = 0, \qquad {\cal B}(x) = \bar {\cal B} (x) = 0, \qquad
R_\mu (x) = \partial_\mu C (x), \nonumber\\
&&\bar R_\mu (x) = \partial_\mu \bar C(x),\qquad
S_\mu =  \partial_\mu B (x), \qquad B (x) + \bar B (x) = 0.
\end{array} \eqno(2.47)
$$
Physically, the condition $\tilde d \tilde A = d A$ amounts to the restriction
that there is no contribution of the superspace variables to the curvature
2-form. Insertions of these auxiliary fields in the above expansion leads to 
$$
\begin{array}{lcl}
B_{\mu} (x, \theta, \bar \theta) &=& A_{\mu} (x) 
+ \theta\; \partial_{\mu} \bar C (x) + \bar \theta\; \partial_{\mu} C (x) 
+ i \;\theta \;\bar \theta \;\partial_{\mu} B (x), \nonumber\\
\Phi (x, \theta, \bar \theta) &=& C (x) 
- i\; \theta \;B (x), \nonumber\\
\bar \Phi (x, \theta, \bar \theta) &=& \bar C (x) 
+ i \; \bar \theta \;B (x), 
\end{array} \eqno(2.48)
$$
which is the same
expansion as quoted in equation (2.44). However, the transformations $s_{(a)b}$
of (2.44) now correspond to the off-shell nilpotent (anti-)BRST transformations 
$\tilde s_{(a)b}$ of (2.4) (i.e. $s_{(a)b} \rightarrow \tilde s_{(a)b}$).
{\it It is interesting to note that the horizontality condition
($\tilde d \tilde A = d A$) enforces the odd superfield $\Phi$ to become
anti-chiral (i.e. $ \partial_{\bar\theta} \Phi = 0$) and the other odd 
superfield $\bar \Phi$ becomes chiral (i.e. $\partial_{\theta} 
\bar \Phi = 0$).}
Similarly, the dual horizontality condition 
($\tilde \delta \tilde A = \delta A$) w.r.t. the
(super) co-exterior derivatives $(\tilde \delta) \delta$ (together with the
definitions in (2.17) and the $\star$ operation in (2.23)), leads to the
following relationships
$$
\begin{array}{lcl}
&&s(x) = \bar s(x) = 0, \qquad B(x) = \bar B (x) = 0, \qquad
R_\mu (x) = - \varepsilon_{\mu\nu} \partial^\nu \bar C (x), \nonumber\\
&&\bar R_\mu (x) = - \varepsilon_{\mu\nu} \partial^\nu  C(x),
\qquad S_\mu =  \varepsilon_{\mu\nu} \partial^\nu {\cal B} (x), 
\qquad {\cal B} (x) + \bar {\cal B} (x) = 0.
\end{array} \eqno(2.49)
$$
Physically, the condition $\tilde \delta \tilde A = \delta A$ amounts to
no superspace contribution to the zero-form gauge-fixing term.
When we substitute the above expressions in the super expansion (2.46), we
get 
$$
\begin{array}{lcl}
B_{\mu} (x, \theta, \bar \theta) &=& A_{\mu} (x) 
- \theta\; \varepsilon_{\mu\nu} \partial^\nu C (x) 
- \bar \theta\; \varepsilon_{\mu\nu} \partial^\nu \bar C (x) 
+ i \;\theta \;\bar \theta \;\varepsilon_{\mu\nu} \partial^\nu {\cal B} (x), 
\nonumber\\
\Phi (x, \theta, \bar \theta) &=& C (x) 
- i \;\bar \theta\; {\cal B} (x) , \nonumber\\
\bar \Phi (x, \theta, \bar \theta) &=& \bar C (x) 
+ i \;\theta\;  {\cal B} (x),
\end{array} \eqno(2.50)
$$
which is in the same form as in (2.45). However, the transformations
$s_{(a)d}$ of (2.45) now correspond to the ones listed in (2.4) for the 
off-shell nilpotent (anti-)co-BRST symmetries $\tilde s_{(a)d}$ 
(i.e. $s_{(a)d} \rightarrow \tilde s_{(a)d}$). 
{\it It is worth pointing out that the dual horizontality condition
$(\tilde \delta \tilde A = \delta A$) enforces the odd $\Phi$ superfield to
become chiral (i.e. $\partial_{\theta} \Phi = 0$) and the other odd superfield
$\bar \Phi$ becomes anti-chiral (i.e. $\partial_{\bar\theta} \bar \Phi = 0$).
Thus, it is obvious that the effect of the horizontality- and dual 
horizontality conditions on the odd superfields is diametrically opposite 
as far as the chirality is concerned.} The geometrical interpretation for
the conserved and off-shell nilpotent charges $Q_{(a)b}$ and $Q_{(a)d}$, as 
the translation generators on the supermanifold,  is the same
as elaborated earlier. The mappings: $\tilde s_{(d)b} \leftrightarrow
\partial/ \partial \bar\theta, \tilde s_{ab} \leftrightarrow \partial/ 
\partial \theta, \tilde s_{ad} \leftrightarrow \partial/\partial \theta$
are valid for the off-shell nilpotent symmetries as well. This can be shown
explicitly by expressing $X$ of (2.8), in terms of superfields, as
$$
\begin{array}{lcl}
\frac{1}{2}\;\tilde s_{b} \tilde s_{ab} \Bigl (i A_\mu A^\mu - \xi \bar C C
\Bigr ) &=& {\displaystyle \frac{1}{2} \frac{\partial}{\partial\bar\theta}
\frac{\partial}{\partial\theta}} \Bigl (i B_\mu B^\mu - \xi \bar\Phi \Phi
\Bigr )|_{\mbox{(anti-)BRST}},\nonumber\\
\tilde s_{b} \Bigl (- i \bar C [ \partial_\mu A^\mu + \frac{\xi}{2} B] \Bigr )
&=& \mbox{Lim}_{\theta,\bar\theta \rightarrow 0}
{\displaystyle \frac{\partial}{\partial\bar\theta }}
\Bigl [- i \bar\Phi (\partial_\mu B^\mu) + \frac{\xi}{2} \bar\Phi
{\displaystyle \frac{\partial\Phi}{\partial \theta} }
\Bigr ]|_{\mbox{(anti-)BRST}}\nonumber\\
&\equiv& 
\mbox{Lim}_{\theta,\bar\theta \rightarrow 0}
{\displaystyle \frac{\partial}{\partial\bar\theta }}
\Bigl [- i \bar\Phi (\partial_\mu B^\mu) - \frac{\xi}{2} \bar\Phi
{\displaystyle \frac{\partial\bar\Phi}{\partial \bar\theta} }
\Bigr ]|_{\mbox{(anti-)BRST}},
\nonumber\\
\tilde s_{ab} \Bigl (i  C [ \partial_\mu A^\mu + \frac{\xi}{2} B] \Bigr )
&=& \mbox{Lim}_{\theta,\bar\theta \rightarrow 0}
{\displaystyle \frac{\partial}{\partial \theta }}
\Bigl [ + i \Phi (\partial_\mu B^\mu) + \frac{\xi}{2} \Phi
{\displaystyle \frac{\partial\bar\Phi}{\partial \bar\theta} }
\Bigr ]|_{\mbox{(anti-)BRST}}
\nonumber\\
&\equiv& 
\mbox{Lim}_{\theta,\bar\theta \rightarrow 0}
{\displaystyle \frac{\partial}{\partial\theta }}
\Bigl [ + i \Phi (\partial_\mu B^\mu) - \frac{\xi}{2} \Phi
{\displaystyle \frac{\partial \Phi}{\partial \theta} }
\Bigr ]|_{\mbox{(anti-)BRST}},
\end{array} \eqno(2.51)
$$
and $Y$ of (2.8) bears an appearance, in terms of superfields,  as
$$
\begin{array}{lcl}
\frac{1}{2}\;\tilde s_{d} \tilde s_{ad} \Bigl (i A_\mu A^\mu - \xi \bar C C
\Bigr ) &=& {\displaystyle \frac{1}{2} \frac{\partial}{\partial\bar\theta}
\frac{\partial}{\partial\theta}} \Bigl (i B_\mu B^\mu - \xi \bar\Phi \Phi
\Bigr )|_{\mbox{(anti-)co-BRST}},\nonumber\\
\tilde s_{d} \Bigl ( i C [ E - \frac{1}{2} {\cal B} ] \Bigr )
&=& \mbox{Lim}_{\theta,\bar\theta \rightarrow 0}
{\displaystyle \frac{\partial}{\partial\bar\theta }}
\Bigl [- i \Phi (\varepsilon^{\mu\nu}\partial_\mu B_\nu) 
+ \frac{1}{2} \Phi
{\displaystyle \frac{\partial\Phi}{\partial \bar\theta} }
\Bigr ]|_{\mbox{(anti-)co-BRST}}\nonumber\\
&\equiv& 
\mbox{Lim}_{\theta,\bar\theta \rightarrow 0}
{\displaystyle \frac{\partial}{\partial\bar\theta }}
\Bigl [- i \Phi (\varepsilon^{\mu\nu}\partial_\mu B_\nu) 
- \frac{1}{2} \bar\Phi
{\displaystyle \frac{\partial\bar\Phi}{\partial \theta} }
\Bigr ]|_{\mbox{(anti-)co-BRST}},
\nonumber\\
\tilde s_{ab} \Bigl (- i \bar C [ E -  \frac{1}{2} {\cal B} ] \Bigr )
&=& \mbox{Lim}_{\theta,\bar\theta \rightarrow 0}
{\displaystyle \frac{\partial}{\partial \theta }}
\Bigl [ + i \bar \Phi (\varepsilon^{\mu\nu} \partial_\mu B_\nu) 
- \frac{1}{2} \bar \Phi
{\displaystyle \frac{\partial \Phi}{\partial \bar\theta} }
\Bigr ]|_{\mbox{(anti-)co-BRST}} \nonumber\\
&\equiv& 
 \mbox{Lim}_{\theta,\bar\theta \rightarrow 0}
{\displaystyle \frac{\partial}{\partial\theta }}
\Bigl [ + i \bar \Phi (\varepsilon^{\mu\nu}\partial_\mu B_\nu) 
+ \frac{1}{2} \bar \Phi
{\displaystyle \frac{\partial \bar \Phi}{\partial \theta} }
\Bigr ]|_{\mbox{(anti-)co-BRST}},
\end{array} \eqno(2.52)
$$
where the above subscripts stand for the expansions 
in (2.48) and (2.50) respectively.\\

\noindent
{\bf 3 Self-interacting 2D non-Abelian gauge theory}\\

\noindent
In this section, the on-shell and off-shell nilpotent (anti-)BRST- 
and (anti-)co-BRST symmetries are discussed in the Lagrangian- and 
superfield formulations.\\

\noindent
{\bf 3.1 (Anti-)BRST and (anti-)co-BRST symmetries: Lagrangian formalism}\\

\noindent
Let us begin with the BRST invariant Lagrangian density ${\cal L}_{b}^{(N)}$
for the self-interacting 2D non-Abelian gauge theory 
\footnote{We follow here the notations such that 
$F_{01} = E = \partial_{0} A_{1} - \partial_{1} A_{0}
+  A_{0} \times  A_{1} \equiv - \varepsilon^{\mu\nu}
(\partial_\mu A_\nu + \frac{1}{2}\; A_\mu \times A_\nu) = F^{10}, 
\varepsilon_{01} = \varepsilon^{10} = + 1, \;D_{\mu} C = \partial_{\mu} C + 
A_{\mu} \times C, \alpha \cdot \beta = \alpha^{a} \beta^{a},
(\alpha \times \beta)^a = f^{abc} \alpha^{b} \beta^{c}$ where $\alpha$
and $\beta$ are the non-null vectors in the group space. 
Here the Greek indices: $\mu, \nu, \rho...= 0, 1$ correspond to the
spacetime directions and Latin indices: 
$ a, b, c...= 1, 2, 3...$ stand for the ``colour'' values in the group space.} 
in an arbitrary gauge (with the gauge parameter $\xi \neq 0$) [43-46]
$$
\begin{array}{lcl}
{\cal L}_{b}^{(N)} &=& - \frac{1}{4}\; F^{\mu\nu}\cdot F_{\mu\nu} 
- \frac{1}{2\xi}\; (\partial_{\mu} A^{\mu}) \cdot (\partial_{\rho} A^{\rho})
- i \;\partial_{\mu} \bar C \cdot D^\mu C, \nonumber\\
&\equiv& \frac{1}{2}\; E \cdot  E 
- \frac{1}{2\xi} \; (\partial_{\mu}  A^{\mu}) \cdot (\partial_{\rho} A^{\rho})
- i \;\partial_{\mu} \bar C \cdot D^\mu C, 
\end{array} \eqno(3.1)
$$
where $F_{\mu\nu} = \partial_{\mu} A_{\nu} -
\partial_{\nu} A_{\mu} + (A_{\mu} \times A_{\nu})$ is the field 
strength tensor derived from the one-form connection $ A = d x^\mu A_\mu \equiv
dx^\mu A^{a}_{\mu} T^{a}$ by the Maurer-Cartan equation $F = d A + A \wedge A$ 
with $ F = \frac{1}{2} d x^\mu \wedge d x^\nu F_{\mu\nu}^{a} T^{a}$. Here
$T^{a}$ are the generators of the compact Lie algebra $ [ T^{a}, T^{b} ]
= f^{abc} T^{c}$ where $f^{abc}$ are the structure constants that can be
chosen to be totally antisymmetric in $ a, b, c$ 
(see, e.g., Ref. [46] for more details). The anti-commuting
($ (C^a)^ 2 = (\bar C^a)^2 = 0, C^a \bar C^b + \bar C^b C^a = 0$) (anti-)ghost
fields $(\bar C^a) C^a$ (which interact with 
the self-interacting non-Abelian gauge fields $A_{\mu}$ only
in the loop diagrams) are required to be present in the theory to maintain the
unitarity and gauge invariance together at any arbitrary order of perturbative
computations [49]. These fields (even though 
they interact with the gauge fields $A_\mu$) are not the physical 
matter fields. The above Lagrangian density (3.1)
respects the following on-shell ($\partial_\mu D^\mu C = D_\mu \partial^\mu
\bar C = 0 )$ nilpotent ($s_{b}^2 = s_{d}^2 = 0$) BRST ($s_{b}$)
and dual(co-)BRST ($s_{d}$) symmetry transformations [31,35]
$$
\begin{array}{lcl}
&& s_{b} A_\mu = D_\mu C, \quad s_{b} C = - \frac{1}{2} C \times C, \quad
s_{b} \bar C = - \frac{i}{\xi} (\partial_\mu A^\mu), \quad s_{b} E 
= E \times C,\nonumber\\ 
&& s_{d} A_\mu = - \varepsilon_{\mu\nu} \partial^\nu \bar C, \qquad 
s_{d} \bar C = 0, \qquad
s_{d} C = - i \; E, \qquad s_{d} E = D_\mu \partial^\mu \bar C. 
\end{array} \eqno(3.2)
$$
It is worth pointing out that the kinetic energy term $\frac{1}{2} (E \cdot E$)
remains invariant under the BRST transformation $s_{b}$. On the other hand, it
is the gauge fixing term ($-\frac{1}{2\xi} (\partial_\mu A^\mu) \cdot 
(\partial_\rho A^\rho)$) that remains unchanged under the dual(co-)BRST
transformations $s_{d}$. The anti-commutator 
$s_{w} = \{ s_{b}, s_{d} \}$ of these nilpotent symmetries leads to the
definition of a bosonic symmetry ($s_{w}^2 \neq 0)$, under which, the ghost
term $- i \partial_\mu \bar C \cdot D^\mu C$ remains invariant [35].
The auxiliary fields
$B$ and ${\cal B}$ can be introduced to linearize the gauge-fixing term
and the kinetic energy term $\frac{1}{2}(E \cdot E)$ 
(because there is no magnetic 
component of $F_{\mu\nu}$ for the $(1+1)$-dimensional (2D) non-Abelian 
gauge theory) as 
$$
\begin{array}{lcl}
{\cal L}_{B}^{(N)} =  {\cal B}\cdot  E - \frac{1}{2} {\cal B} \cdot {\cal B}
+ B \cdot (\partial_{\mu}  A^{\mu}) + \frac{\xi}{2}\; B \cdot B
- i \partial_{\mu} \bar C \cdot D^\mu C. 
\end{array} \eqno(3.3)
$$
The above Lagrangian density
(3.3) respects the following off-shell nilpotent
$(\tilde s_{b}^2 = 0,  \tilde s_{d}^2 = 0)$ BRST ($\tilde s_{b}$)
-and dual(co)-BRST ($\tilde s_{d}$) symmetry transformations [31,35,39]
$$
\begin{array}{lcl}
\tilde s_{b} A_{\mu} &=& D_{\mu} C, \qquad 
\tilde s_{b} C = -\frac{1}{2} C \times C, \qquad 
\tilde s_{b} \bar C = i B,  \nonumber\\
\tilde s_{b} {\cal B} &=& {\cal B} \times C, \qquad \;\; 
\tilde s_{b} B = 0, \;\;\qquad
\;\tilde s_{b} E = E \times C,
\end{array}\eqno(3.4)
$$
$$
\begin{array}{lcl}
\tilde s_{d} A_{\mu} &=& - \varepsilon_{\mu\nu} \partial^\nu \bar C, \qquad
\tilde s_{d} \bar C = 0, \qquad \tilde s_{d} C = - i {\cal B}, \nonumber\\
\tilde s_{d} {\cal B} &=& 0, \;\;\;\;\qquad \tilde s_{d} B = 0, 
\;\;\;\;\qquad \tilde s_{d} E = D_{\mu} \partial^{\mu} \bar C.
\end{array}\eqno(3.5)
$$
Besides BRST- and co-BRST symmetry transformations (3.4) and (3.5), there
are anti-BRST- and anti-co-BRST symmetries that are also
present in the theory. To realize
these, one has to introduce another auxiliary field $\bar B$ 
(satisfying $ B + \bar B = i\; C \times \bar C$) to recast the
Lagrangian density (3.3) into the following forms [50]
$$
\begin{array}{lcl}
{\cal L}_{\bar B}^{(N)} 
=  {\cal B}\cdot  E - \frac{1}{2} {\cal B} \cdot {\cal B}
+ B \cdot (\partial_{\mu}  A^{\mu}) + \frac{\xi}{2} (B \cdot B + \bar B \cdot
\bar B) - i \partial_{\mu} \bar C \cdot D^\mu C, 
\end{array} \eqno(3.6a)
$$
$$
\begin{array}{lcl}
{\cal L}_{\bar B}^{(N)} 
=  {\cal B}\cdot  E - \frac{1}{2} {\cal B} \cdot {\cal B}
- \bar B \cdot (\partial_{\mu}  A^{\mu}) + \frac{\xi}{2} (B \cdot B + \bar B 
\cdot \bar B) - i D_{\mu} \bar C \cdot \partial^\mu C.
\end{array} \eqno(3.6b)
$$
The Lagrangian density (3.6b) transforms to a total derivative under the 
following off-shell nilpotent 
anti-BRST (${\tilde s}_{ab}$)- and (anti-)co-BRST (${\tilde s}_{ad}$) symmetry
transformations [35,39]
$$
\begin{array}{lcl}
{\tilde s}_{ab} A_{\mu} &=& D_{\mu} \bar C, \qquad \;\;{\tilde s}_{ab} \bar C 
= - \frac{1}{2} \bar C \times \bar C, \qquad \;\;
\tilde s_{ab}  C = i \bar B, \qquad \;\;\tilde s_{ab} \bar B = 0, \nonumber\\
\tilde s_{ab} E &=& E \times \bar C, \quad  \tilde s_{ab} {\cal B} =
{\cal B} \times \bar C, \quad \tilde s_{ab} B = B \times \bar C,
\quad \tilde s_{ab} (\partial_{\mu} A^\mu) = \partial_{\mu} D^\mu \bar C,
\end{array} \eqno(3.7)
$$
$$
\begin{array}{lcl}
\tilde s_{ad} A_{\mu} &=& - \varepsilon_{\mu\nu} \partial^\nu C, \qquad
\tilde s_{ad}  C = 0, \qquad \tilde s_{ad} \bar C = + i {\cal B}, \qquad
\tilde s_{ad}  {\cal B} = 0, \nonumber\\
\tilde s_{ad} E &=& D_{\mu} \partial^\mu C, \qquad \tilde s_{ad} \bar B = 0,
\qquad \tilde s_{d} B = 0, \qquad \tilde s_{ad} (\partial_\mu A^\mu) = 0.
\end{array} \eqno(3.8)
$$
All the above continuous symmetry
transformations can be concisely expressed, in terms of the Noether
conserved charges $Q_{r}$ as quoted in (2.5).

The Lagrangian density (3.1) can be expressed, modulo
some total derivatives,  as the sum of terms that
turn out to be (co-)BRST anti-commutators
$$
\begin{array}{lcl}
{\cal L}_{b}^{(N)} = \{ Q_{d}, \frac{1}{2} E \cdot C \} - \{ Q_{b}, 
\frac{1}{2} (\partial_\mu A^\mu) \cdot \bar C \} \equiv
s_{d} \bigl [\; \frac{i}{2} E \cdot C\; \bigr ]
- s_{b} \bigl [\; \frac{i}{2} (\partial_\mu A^\mu) \cdot \bar C\; \bigr ],
\end{array} \eqno(3.9)
$$
which resembles very much like the Lagrangian density for the Witten type
TFT if we assume that the vacuum and physical states of this theory are
annihilated by the (co-) BRST charges (i.e. $Q_{(d)b} |phys> = 0,
Q_{(d)b} |vac> = 0$). Such a situation does arise when we choose the
harmonic state of the Hodge decomposed state to be the
physical state (including the vacuum) in the quantum Hilbert space of states.
This choice is guided by some aesthetic reasons because the harmonic states 
possess the maximum symmetries as they remain invariant under (co-)BRST 
symmetries as well as a bosonic symmetry (that is an anti-commutator of 
nilpotent (co-)BRST symmetries). In contrast to the appearance of the
Lagrangian density (which is like Witten type TFT), the local symmetries
of the theory are just like Schwarz type TFT because there are no topological
{\it shift symmetries} in the theory (see, e.g., Ref. [5] for details). 
The Lagrangian density in (3.3) can be also expressed as
$$
\begin{array}{lcl}
{\cal L}_{B}^{(N)} = {\cal B} \cdot E - \frac{1}{2} {\cal B} \cdot {\cal B}
+ \tilde s_{b} \bigl ( - i \bar C \cdot 
[ (\partial_\mu A^\mu) + \frac{\xi}{2} B ] \bigr ).
\end{array} \eqno(3.10)
$$
Similarly, the Lagrangian densities in (3.6) can be re-written as
$$
\begin{array}{lcl}
{\cal L}_{\bar B}^{(N)} &=& {\cal B} \cdot E 
- \frac{1}{2} {\cal B} \cdot {\cal B}
+ \tilde s_{b} \tilde s_{ab} \bigl (\; \frac{i}{2} A_\mu \cdot A^\mu
- \xi\; \bar C \cdot C \; \bigr ),\nonumber\\
{\cal L}_{\bar B}^{(N)} &=& \frac{\xi}{2} (B \cdot B + \bar B \cdot \bar B)
+ B \cdot (\partial_\mu A^\mu) + \tilde s_{d} \bigl ( + i\; C \cdot
[ E - \frac{1}{2} {\cal B} ] \bigr ), \nonumber\\
{\cal L}_{\bar B}^{(N)} &=& \frac{\xi}{2} (B \cdot B + \bar B \cdot \bar B)
- \bar B \cdot (\partial_\mu A^\mu) + \tilde s_{ad} \bigl (- i\; \bar C \cdot
[ E - \frac{1}{2} {\cal B} ] \bigr ). 
\end{array} \eqno(3.11)
$$
The above forms of the Lagrangian densities are just the analogues of 
the corresponding forms for the Abelian gauge theories in (2.8). In fact,
modulo some total derivatives, the expression
$\tilde s_{b} \tilde s_{ab} (\frac{1}{2} A_\mu \cdot A^\mu 
- \xi \bar C\cdot C)
= B \cdot (\partial_\mu A^\mu) - i \partial_\mu \bar C \cdot D^\mu C$. 
However, it can  be shown that the same expression is also equivalent to:
$ - \bar B \cdot (\partial_\mu A^\mu) - i D_\mu \bar C \cdot \partial^\mu C
+ \partial_\mu X^\mu$ (with $X^\mu = - i\; \bar C \cdot A^\mu \times C$) by
exploiting the relationship $ B + \bar B = i (C \times \bar C) $ [50]. 
To confirm 
the topological nature of the above theory, it can be seen that the symmetric
energy momentum for the Lagrangian density (3.1) is [35,39]
$$
\begin{array}{lcl}
T_{\alpha\beta}^{(N)} &=& - \frac{1}{2 \xi}\; (\partial_\rho A^\rho)\;\cdot
[\; \partial_\alpha A_\beta + \partial_\beta A_\alpha \;] - \frac{1}{2} \; E
\cdot
\;[\; \varepsilon_{\alpha\rho} \partial_\beta A^\rho 
+ \varepsilon_{\beta\rho} \partial_\alpha A^\rho \;] \nonumber\\
&-& \frac{i} {2}\; \partial_\alpha \bar C \cdot (\partial_\beta C + D_\beta C)
- \frac{i}{2} \; \partial_\beta \bar C \cdot (\partial_\alpha C + D_\alpha C)
- \eta_{\alpha\beta} {\cal L}_{b}^{(N)},  
\end{array} \eqno(3.12)
$$
which turns out to be the sum of (co-)BRST anti-commutators as given below
$$
\begin{array}{lcl}
T_{\alpha\beta}^{(N)} &=& \{ Q_{b}, L_{\alpha\beta}^{(1)} \}
+ \{ Q_{d}, L_{\alpha\beta}^{(2)} \} \equiv
s_{b} [\; i L_{\alpha\beta}^{(1)} \;]
+ s_{d} [\; i L_{\alpha\beta}^{(2)} \;], 
\nonumber\\
L_{\alpha\beta}^{(1)} &=& \frac{1}{2} \;\bigl  [ 
\;(\partial_\alpha \bar C) \cdot A_\beta +
(\partial_\beta \bar C) \cdot A_\alpha 
+ \eta_{\alpha\beta} (\partial_\rho A^\rho)
\cdot \bar C \;\bigr ], \nonumber\\
L_{\alpha\beta}^{(2)} &=& \frac{1}{2} \;\bigl  [ 
\;(\partial_\alpha C) \cdot \varepsilon_{\beta\rho} A^\rho +
(\partial_\beta C) \cdot  \varepsilon_{\alpha\rho} A^\rho 
- \eta_{\alpha\beta} E \cdot C \;\bigr ].
\end{array} \eqno(3.13)
$$
The form of the above symmetric energy momentum tensor for the non-Abelian
gauge theory ensures that the VEV of the energy density
$<vac | T_{00}^{(N)} |vac> = 0$ is zero and there are no energy excitation
in the physical sector of the theory (i.e. $<phys | T_{00}^{(N)} |phys^\prime>
= 0$) because of the fact that the harmonic (physical) states satisfy:
$Q_{d(b)} |vac> = 0, Q_{d(b)} |phys> = 0$
w.r.t. the (co-)BRST charges $Q_{(d)b}$ {\it which are conserved, nilpotent,
metric independent and hermitian}. In fact, there
are four conserved and nilpotent ($Q_{(a)b}^2 = 0, Q_{(a)d}^2 = 0$)
charges in the theory. As a consequence, there are four sets of topological 
invariants which obey a specific kind of recursion relations. On a 2D
compact manifold, these invariants have been computed [35] in the same
generic form as illustrated in (2.11). The most important quantities in this
connection are the explicitly (anti-)BRST- and (anti-)co-BRST invariant
{\it zero-forms} because, all the rest of the higher degree forms, can be
computed from them by exploiting the recursion relations. For the on-shell
($\partial_\mu D^\mu C = D_\mu \partial^\mu \bar C = 0$) nilpotent (co-)BRST
charges, these zero forms $(W_{0})V_{0}$ are [31,35]
$$
\begin{array}{lcl}
W_{0} = E \cdot \bar C, \qquad V_{0} = - \frac{1}{\xi}
(\partial_\mu A^\mu) \cdot C - \frac{i}{2}\; \bar C \cdot (C \times C).
\end{array} \eqno(3.14)
$$
The analogous zero-forms $A_{0}^{(ab)}, B_{0}^{(b)}$ and
$C_{0}^{(ad)}, D_{0}^{(d)}$ with respect to the off-shell nilpotent 
(anti-)BRST ($Q_{(a)b}$) and
(anti-)co-BRST ($Q_{(a)d}$) charges are
$$
\begin{array}{lcl}
A_{0}^{(ab)} &=& -  \bar B \cdot \bar C + \frac{i}{2}\; C \cdot 
(\bar C \times \bar C), \qquad
C_{0}^{(ad)} = {\cal B} \cdot C, \nonumber\\
A_{0}^{(b)} &=& B \cdot C - \frac{i}{2}\; \bar C \cdot 
(C \times C), \qquad
D_{0}^{(d)} = {\cal B} \cdot \bar C.
\end{array} \eqno(3.15)
$$

\noindent
{\bf 3.2 On-shell nilpotent (co-)BRST symmetries: chiral superfields}\\

\noindent
In this subsection, we shall discuss some of the key features of our work
in [41] for the derivation of the on-shell nilpotent (co-)BRST
symmetries (3.2) in the framework of chiral superfield formulation. First of
all, we generalize the generic local field $\Psi = (A_\mu, C, \bar C)$
(for the basic local fields of the Lagrangian density (3.1)) to a chiral
($\partial_{\theta} \tilde A_{M} = 0$)
supervector superfield $\tilde A_{M} (x,\bar \theta)
= (B_\mu, \Phi, \bar \Phi) (x,\bar \theta)$ with the following super expansion
for all the component superfields along the $\bar\theta$-direction of the
chiral supermanifold
$$
\begin{array}{lcl}
(B_{\mu}^a T^a) (x,  \bar \theta) &=& (A_{\mu}^a T^a) (x) 
+\; \bar \theta\; (R_{\mu}^a T^a) (x), \nonumber\\ 
(\Phi^a T^a) (x,  \bar \theta) &=& (C^a T^a) (x) 
- i\; \bar \theta \;({\cal B}^a T^a) (x), \nonumber\\
(\bar \Phi^a T^a) (x,  \bar \theta) &=& (\bar C^a T^a) (x) 
+ i\; \bar \theta \;(B^a T^a) (x), 
\end{array} \eqno(3.16)
$$
which is the chiral limit ($\theta \rightarrow 0$) of the super expansion 
(2.46) (or (2.14))
on the $(2+2)$-dimensional supermanifold. Note that we have taken here a
minus sign in the expansion of $\Phi = \Phi^aT^a$ {\it only for the algebraic
convenience.} The horizontality condition ($\tilde F = \tilde D \tilde A
= D A = F$) with the following definitions on the chiral supermanifold
$$
\begin{array}{lcl}
\tilde d &=& d Z^M \;\partial_{M} \equiv
d x^\mu \;\partial_\mu + d \bar \theta\;\partial_{\bar\theta}, \quad
\tilde D = \tilde d + \tilde A, \quad D = d + A,
\nonumber\\
\tilde  A &=& d Z^M\; \tilde (A_{M}^a T^a) \equiv
dx^\mu \;(B_\mu^a T^a) (x,\bar\theta) 
+ d \bar \theta\; (\Phi^a T^a) (x,\bar\theta),
\end{array} \eqno(3.17a)
$$
provides a specific kind of
relationship between the auxiliary fields and the basic fields of
the Lagrangian density (3.1) as given below [41]
$$
\begin{array}{lcl}
R_\mu (x) = D_\mu C (x), \qquad {\cal B} (x) = - \frac{i}{2} (C \times C) (x),
\qquad [ {\cal B}, C ] = 0 \rightarrow {\cal B} \times C = 0. 
\end{array} \eqno(3.17b)
$$
The above horizontality
condition does not fix the auxiliary field $ B(x) = (B^a T^a)(x)$ in terms
of the basic fields of (3.1). However, the equation of motion 
($ B (x) = - \frac{1}{\xi} (\partial_\mu A^\mu) (x)$) for the
Lagrangian density (3.3) comes to our rescue. Now the expansion (3.16), in
the concise notation $B_\mu (x, \bar \theta) = (B_\mu^a T^a) (x, \bar\theta)$
etc., becomes
$$
\begin{array}{lcl}
B_{\mu} (x,  \bar \theta) &=& A_{\mu} (x) 
+\; \bar \theta\; D_{\mu} C (x)
\equiv A_\mu (x) + \bar\theta\; (s_{b} A_\mu (x)), \nonumber\\ 
\Phi (x,  \bar \theta) &=& C (x) 
- \; \bar \theta \;\frac{1}{2} (C \times C) (x)
\equiv C (x) + \bar\theta\; (s_{b} C (x)), \nonumber\\
\bar \Phi (x,  \bar \theta) &=& \bar C (x) 
- \frac{i}{\xi} \; \bar \theta \;(\partial_\mu A^\mu) (x)
\equiv \bar C (x) + \bar\theta\; (s_{b} \bar C(x)). 
\end{array} \eqno(3.18)
$$
We note here, vis-a-vis equation (2.5), the following interesting relationship
$$
\begin{array}{lcl}
{\displaystyle \frac{\partial}{\partial \bar \theta}}\; 
\tilde A_{M} (x,\bar\theta)
= - i \;\bigl [ \Psi (x), Q_{b} \bigr ]_{\pm} \equiv s_{b} \Psi (x), \quad
\tilde A_{M} = (\Phi, \bar \Phi, B_\mu), \quad \Psi = (C , \bar C, A_\mu),
\end{array} \eqno(3.19)
$$
which establishes the fact that $Q_{b}$ is the generator of translation
($\partial / \partial\bar\theta$) 
along the Grassmannian direction $\bar\theta$ 
of the supermanifold for the superfield $\tilde A_{M} (x,\bar\theta)$.
This process of translation produces an internal transformation 
$(s_{b} \Psi (x))$ for the local generic field. Hence there is a 
mapping $s_{b} \leftrightarrow \partial / \partial\bar\theta$ which shows
that there is a deep inter-relationship between the {\it internal 
transformation} on the local field $\Psi$ and the translation of the 
superfield $\tilde A_{M}$ along the $\bar\theta$-direction of the
supermanifold. Furthermore, the nilpoency of $s_{b}^2 = 0$ is
connected with $(\partial/\partial\bar\theta)^2 = 0$. In a similar fashion,
the dual horizontality condition ($\tilde \delta \tilde A 
= \delta A$) (with $\tilde \delta = - \star \tilde d \star,
\delta = - * d * $) leads to
$$
\begin{array}{lcl}
\partial_\mu B^\mu + s^{\bar\theta\bar\theta} (\partial_{\bar\theta} \bar \Phi)
- \varepsilon^{\mu\bar\theta} (\partial_\mu \bar \Phi + \varepsilon_{\mu\nu}
\partial_{\bar\theta} B^\nu) = \partial_\mu A^\mu,
\end{array} \eqno(3.20)
$$
where $s^{\bar\theta\bar\theta}$ and $\varepsilon^{\mu\bar\theta}$ are defined
in equation (2.23). The above equality yields
$$
\begin{array}{lcl}
R_\mu (x) = - \varepsilon_{\mu\nu} \partial^\nu \bar C (x), \quad
B (x) = 0, \quad \partial_\mu B^\mu (x,\theta) 
= \partial_\mu A^\mu (x) \rightarrow
\partial_\mu R^\mu (x) = 0.
\end{array} \eqno(3.21)
$$
It is evident that the condition $\partial_\mu R^\mu (x) = 0$ is satisfied
automatically with $R_\mu = - \varepsilon_{\mu\nu} \partial^\nu \bar C$.
The above restriction $\tilde \delta \tilde A = \delta A$ does not fix
the auxiliary field ${\cal B} (x)$ in terms of the basic local fields
$\Psi (x)$. However, the equation of motion $ {\cal B} (x) = E (x)$ for
the Lagrangian density (3.3) comes in handy. With the above
values, the super expansion (3.16) now becomes
$$
\begin{array}{lcl}
B_{\mu} (x, \bar \theta) &=& A_{\mu} (x) 
-\; \bar \theta\; \varepsilon_{\mu\nu} \partial^{\nu} \bar C (x)
\equiv A_\mu (x) + \bar \theta\; (s_{d} A_\mu (x)), \nonumber\\ 
\Phi (x,  \bar \theta) &=& C (x) 
- i \bar \theta\; E (x)
\equiv C (x) + \bar \theta\; (s_{d} C (x)), \nonumber\\
\bar \Phi (x, \bar \theta) &=& \bar C (x) 
+ i\; \bar \theta \;(B (x) = 0)
\equiv \bar C (x) + \theta \; (s_{d} \bar C (x) = 0). 
\end{array} \eqno(3.22)
$$
This super expansion implies the following
$$
\begin{array}{lcl}
{\displaystyle \frac{\partial}{\partial \bar \theta}}\; 
\tilde A_{M} (x,\bar\theta)
= - i \;\bigl [ \Psi (x), Q_{d} \bigr ]_{\pm} \equiv s_{d} \Psi (x), \quad
\tilde A_{M} = (\Phi, \bar \Phi, B_\mu), \quad \Psi = (C , \bar C, A_\mu),
\end{array} \eqno(3.23)
$$
which shows that $Q_{d}$ {\it  is equivalent to a translation generator
$\partial / \partial\bar\theta$ along the 
$\bar\theta$-direction of the supermanifold.}
 The process of the translation of the superfield $\tilde A_{M} (x,\bar\theta)$
along the $\bar\theta$-direction leads to the co-BRST transformation
$s_{d} \Psi$ on the local field $\Psi = (A_\mu, C, \bar C)$. Thus, we have
a mapping $s_{d} \leftrightarrow \partial / \partial\bar\theta$ which
establishes an inter-relationship between the {\it internal transformation}
$s_{d}$ on the local field $\Psi$ and the {\it translation} of the superfield
$\tilde A_{M}$ along the $\bar\theta$-direction of the supermanifold. In 
addition, the nilotency $s_{d}^2 = 0$ property of the
internal transformation $s_{d}$ on the local field $\Psi$ is connected with
the property of the square of the translation generator 
$(\partial/\partial\bar\theta)^2 = 0$ (i.e. a couple of successive
translations) of the supervector superfield $\tilde A_{M}$
along $\bar\theta$-direction of the supermanifold. {\it Even  though
$s_{(d)b} \leftrightarrow \partial/\partial\bar\theta$, 
there is a clear distinction
between them. Under $s_{b}$, both the (anti-)ghost fields 
$(\bar C)C$ transform. However, under $s_{d}$ only the ghost field $C$
transforms but the anti-ghost fields remains intact.} In more sophisticated 
language, it can be seen that the superfields $\Phi$ and $\bar\Phi$ remain
{\it chiral} superfield under transformations generated by $s_{b}$. However,
under transformations generated by $s_{d}$, only the superfield $\Phi$
remains chiral but $\bar \Phi$ becomes {\it only a 
spacetime dependent local field}. In other words, the local anti-ghost field
$\bar C (x) $ does not transform under $s_{d}$.\\

\noindent
{\bf 3.3 Anti-chiral superfield formalism: no symmetries}\\

\noindent
Unlike the case of Abelian gauge theory, we show here that 
the choice of the anti-chiral
superfields ($\partial_{\bar \theta} \tilde A_{M} = 0$)
does not lead to any important symmetries in spite of the fact
that we exploit the (dual) horizontality conditions.
The super expansions for the component superfields of the supervector 
anti-chiral superfield $\tilde A_M (x,\theta) 
= (B_\mu, \Phi, \bar \Phi) (x,\theta)$ along the $\theta$-direction of
the chiral supermanifold are
$$
\begin{array}{lcl}
(B_{\mu}^a T^a) (x, \theta) &=& (A_{\mu}^a T^a) (x) 
+ \theta\; (\bar R_{\mu}^a T^a) (x), \nonumber\\
(\Phi^a T^a) (x, \theta) &=& (C^aT^a) (x) 
- i\; \theta  (B^aT^a) (x), \nonumber\\
(\bar \Phi^aT^a) (x, \theta) &=& (\bar C^aT^a) (x) 
+ i \;\theta\; ({\cal B}^aT^a) (x), 
\end{array} \eqno(3.24)
$$
which is the anti-chiral limit ($\bar\theta \rightarrow 0$) of the expansion
on $(2+2)$-dimensional supermanifold in (2.14) with the group valued 
basic- as well as auxiliary fields. We exploit now the horizontality condition
($\tilde F = \tilde D \tilde A = D A = F$) with the following definitions on
the anti-chiral supermanifold 
$$
\begin{array}{lcl}
\tilde d &=& d Z^M \;\partial_{M} \equiv
d x^\mu \;\partial_\mu + d  \theta\;\partial_{\theta}, \quad
\tilde D = \tilde d + \tilde A, \quad D = d + A,
\nonumber\\
\tilde  A &=& d Z^M\; (\tilde A_{M}^a T^a) \equiv
dx^\mu \;(B_\mu^a T^a) (x,\theta) 
+ d \theta\; (\bar \Phi^a T^a) (x,\theta).
\end{array} \eqno(3.25)
$$
The explicit expressions for the individual terms in $\tilde D \tilde A
= \tilde d \tilde A + \tilde A \wedge \tilde A$ are
$$
\begin{array}{lcl}
\tilde d \tilde A &=& (dx^\mu \wedge dx^\nu) (\partial_\mu B_\nu)
+ (dx^\mu \wedge d\theta) (\partial_\mu \bar\Phi - \partial_\theta B_\mu)
- (d\theta \wedge d \theta) (\partial_\theta \bar\Phi), \nonumber\\
\tilde A \wedge \tilde A &=& (dx^\mu \wedge dx^\nu) (B_\mu B_\nu)
+ (dx^\mu \wedge d\theta) ( [ B_\mu, \bar\Phi ] )
- (d\theta \wedge d \theta) (\bar \Phi \bar\Phi).
\end{array} \eqno(3.26)
$$
The horizontality restrictions result in the following relationships
$$
\begin{array}{lcl}
&&\partial_{\theta} \bar \Phi + \frac{1}{2} \{ \bar \Phi, \bar \Phi \} = 0,
\quad \partial_\mu \bar \Phi - \partial_\theta B_\mu + [B_\mu, \bar \Phi] = 0,
\nonumber\\
&& \tilde F_{\mu\nu} \equiv 
\partial_\mu B_\nu - \partial_\nu B_\mu + [ B_\mu, B_\nu ] =
\partial_\mu A_\nu - \partial_\nu A_\mu + [ A_\mu, A_\nu ] \equiv F_{\mu\nu},
\end{array} \eqno(3.27)
$$
which lead to
$$
\begin{array}{lcl}
\bar R_\mu (x) = D_\mu \bar C (x), \qquad {\cal B} (x) 
=  \frac{i}{2} (\bar C \times \bar C) (x),
\qquad [ {\cal B}, \bar C ] = 0 \rightarrow {\cal B} \times \bar C = 0. 
\end{array} \eqno(3.28)
$$
The l.h.s. of the last relationship in (3.27) becomes
$\tilde F_{\mu\nu} = F_{\mu\nu} + \theta [D_\mu, D_{\nu}] \bar C
= F_{\mu\nu} + \theta \; F_{\mu\nu} \times \bar C$ which is consistent, 
in the sense that, the kinetic energy term of the Lagrangian density
(3.1) remains invariant (i.e. $- \frac{1}{4} \tilde F_{\mu\nu} \cdot \tilde
F^{\mu\nu} = - \frac{1}{4} F_{\mu\nu} \cdot F^{\mu\nu}$). The horizontality
condition does not fix $B(x)$ in terms of the basic fields of the 
Lagrangian density (3.1). The equation of motion for the Lagrangian density
(3.3), however, yields a connection $B(x) = - \frac{1}{\xi} 
(\partial_\mu A^\mu) (x)$. Thus, expansion (3.24) becomes 
$$
\begin{array}{lcl}
B_{\mu} (x, \theta) &=& A_{\mu} (x) 
+ \theta\; D_{\mu} \bar C (x)
\equiv A_\mu (x) + \theta \; (\bar s_{b} A_\mu (x)), \nonumber\\
\Phi (x, \theta) &=& C (x) 
+ \frac{i}{\xi}\; \theta  (\partial_\mu A^\mu) (x)
\equiv C (x) + \theta\; (\bar s_{b} C (x)), \nonumber\\
\bar \Phi (x, \theta) &=& \bar C (x) 
- \frac{1}{2}\;\theta\; (\bar C \times \bar C) (x)
\equiv \bar C (x) + \theta\; (\bar s_{b} \bar C (x)), 
\end{array} \eqno(3.29)
$$
for some transformation $\bar s_{b}$ (which correspond to translation
along $\theta$-direction of the anti-chiral supermanifold). However, there 
are a few problems. First, it can be seen that $\bar s_{b}$ is not the
symmetry transformation for (3.1) (i.e. $\bar s_{b} {\cal L}_{b}^{(N)} 
\neq 0, \bar s_{b} {\cal L}_{b}^{(N)} \neq \partial_\mu X^\mu$ where
$X^\mu$ is in terms of some local fields). Second,
it is not a nilpotent transformation (e.g., $\bar s_{b}^2 C \sim
\partial_\mu D^\mu \bar C \neq 0$
because the equation of motion for (3.1) is
$D_\mu \partial^\mu \bar C = 0$
(and $\partial_\mu D^\mu \bar C \neq 0$)). Finally, the interpretation in 
terms of the translation along Grassmannian 
$\theta$-direction becomes problematic because of the fact that,
even though $(\partial/\partial \theta)^2 = 0$, the internal
transformation $\bar s_{b}$ under consideration is {\it not}
a nilpotent transformation (i.e. $\bar s_{b}^2 \neq 0$). Similarly,
it can be checked that, for the application of dual horizontality condition
($\tilde \delta \tilde A = \delta A$), we have the following expressions
$$
\begin{array}{lcl}
\star\; \tilde A &=& \varepsilon^{\mu\nu} (dx_\nu) B_\mu (x,\theta)
+ (d \theta)\; \Phi (x, \theta), \nonumber\\
\tilde \delta \tilde A &=& - \star \tilde d \star \tilde A
= \partial_\mu B^\mu + s^{\theta\theta} (\partial_{\theta} \Phi)
- \varepsilon^{\mu\theta} (\partial_\mu \Phi + \varepsilon_{\mu\nu}
\partial_{\theta} B^\nu), \nonumber\\
\delta A &=& -\; *\; d\; *\;  A = \partial_\mu A^\mu,
\end{array} \eqno(3.30)
$$
where $s^{\theta\theta}$ and $\varepsilon^{\mu\theta}$ are defined in
equation (2.23). The equality $\tilde \delta \tilde A = \delta A$ 
produces the following relationships between auxiliary fields and
basic fields of (3.1)
$$
\begin{array}{lcl}
\partial_{\theta} \Phi = 0 \rightarrow B (x) = 0, \qquad
\partial_\mu \Phi + \varepsilon_{\mu\nu} \partial_\theta B^\nu = 0
\rightarrow \bar R_\mu = - \varepsilon_{\mu\nu} \partial^\nu  C.
\end{array} \eqno(3.31)
$$
The additional restriction $\partial_\mu B^\mu = \partial_\mu A^\mu$ implies 
$\partial_\mu \bar R^\mu = 0$ which is trivially satisfied by
$\bar R_\mu = - \varepsilon_{\mu\nu} \partial^\nu C$. The above dual
horizontality condition does not fix ${\cal B} (x)$ in terms of the basic
fields of (3.1). However, for the Lagrangian density (3.3), the equation
of motion is: ${\cal B} = E$. Thus, the expansion in (3.24) can be
expressed in terms of a transformation $\bar s_{d}$ as
$$
\begin{array}{lcl}
B_{\mu} (x, \theta) &=& A_{\mu} (x) 
- \theta\; \varepsilon_{\mu\nu} \partial^\nu  C (x)
\equiv A_\mu (x) + \theta \; (\bar s_{d} A_\mu (x)), \nonumber\\
\Phi (x, \theta) &=& C (x) 
- i\;\theta  (B (x) = 0)
\equiv C (x) + \theta\; (\bar s_{d} C (x) = 0), \nonumber\\
\bar \Phi (x, \theta) &=& \bar C (x) 
+ i\;\theta\; E (x)
\equiv \bar C (x) + \theta\; (\bar s_{d} \bar C (x)).
\end{array} \eqno(3.32)
$$
However, as it turns out, the transformations $\bar s_{d}$ are not the
symmetry transformation for the Lagrangian density (3.1)
(i.e. $\bar s_{d} {\cal L}_{b}^{(N)} \neq 0 , \bar s_{d} {\cal L}_{b}^{(N)}
\neq \partial_\mu Y^\mu$ where $Y^\mu$ stands for 
an expression in terms of some local fields). 
Furthermore, the transformation $\bar s_{d}$ is
not an on-shell  nilpotent symmetry transformation as is evident from
$\bar s_{d}^2 \bar C \sim D_\mu \partial^\mu C \neq 0$. 
In fact, the equation of motion resulting from the Lagrangian density (3.1) 
is $\partial_\mu D^\mu C = 0$ (and $D_\mu \partial^\mu C \neq 0$). 
We conclude, therefore,
that {\it the on-shell nilpotent} anti-BRST and anti-co-BRST symmetries
do not exist for any of the Lagrangian densities
 quoted above for the 2D non-Abelian
gauge theory. The same conclusion emerges even if we start with the
anti-chiral limit of the expansion in (2.46). Thus, the exercise performed
in this subsection provides, in some sense, a concrete proof of the
reason behind the non-existence of the on-shell nilpotent version of the
anti-BRST and anti-co-BRST symmetries for the 2D non-Abelian gauge theory. 
In fact, to the best of our knowledge, such 
on-shell nilpotent symmetries do not exist for
the non-Abelian gauge theories in any arbitrary dimension of spacetime.
As discussed in subsection 3.1, only the off-shell nilpotent version of the
anti-BRST and anti-co-BRST symmetries exist for the self-interacting
non-Abelian gauge theory in two-dimensions of spacetime.\\ 

\noindent
{\bf 3.4 Off-shell nilpotent (anti-)BRST and (anti-)co-BRST symmetries: 
general superfield formulation}\\

\noindent
We shall capture in this subsection  some of the key points of our work [39]
where the off-shell nilpotent (anti-)BRST and (anti-)co-BRST symmetries
have been found in superfield formalism. We start off with the super
expansion (2.46) but all the superfields $(\tilde A_{M}^aT^a) 
(x,\theta,\bar\theta)
= (B_\mu^aT^a, \Phi^aT^a, \bar \Phi^aT^a) (x,\theta,\bar\theta)$
as well as local fields (e.g. $A_\mu = A_\mu^aT^a, C = C^aT^a$ etc.)
are group valued. It should be also noticed that the degrees of freedom
of the fermionic (odd) fields $R_\mu, \bar R_\mu, C, \bar C, s, \bar s$
match with that of the bosonic (even) fields $A_\mu, S_\mu, B, \bar B,
{\cal B}, \bar {\cal B}$ so that the theory can be consistent with the
basic requirements of supersymmetry. The horizontality restriction
$\tilde F = \tilde D \tilde A = D A = F$ (where $\tilde D \tilde A
= \tilde d \tilde A + \tilde A \wedge \tilde A,
D A = d A + A \wedge A$) leads to the following
relationships [39]
$$
\begin{array}{lcl}
R_\mu (x) &=& D_\mu C (x), \quad \bar R_\mu (x) = D_\mu \bar C (x), 
\quad B (x) + \bar B (x)  = i (C \times \bar C) (x), \nonumber\\
{\cal B} (x) &=& - \frac{i}{2} (C \times C) (x), \quad 
\bar {\cal B} (x) = - \frac{i}{2} (\bar C \times \bar C) (x), \quad 
\bar s(x) = - (B \times \bar C) (x), \nonumber\\
S_\mu (x) &=& D_\mu B (x) - (D_\mu C \times \bar C) (x), 
\qquad s(x) = (\bar B \times C) (x), 
\end{array} \eqno(3.33)
$$
where $S_\mu (x)$ can be equivalently written as:
$S_\mu (x) = - D_\mu \bar B (x) - (D_\mu \bar C \times C) (x)$ and 
the individual terms in $\tilde F = \tilde d \tilde A + \tilde A \wedge
\tilde A$ have been computed as
$$
\begin{array}{lcl}
\tilde d \tilde A &=& (dx^\mu \wedge dx^\nu) (\partial_\mu B_\nu)
+ (dx^\mu \wedge d\theta) (\partial_\mu \bar\Phi - \partial_\theta B_\mu)
- (d\theta \wedge d \theta) (\partial_\theta \bar\Phi) \nonumber\\
&+& (dx^\mu \wedge d \bar\theta) 
(\partial_\mu \Phi - \partial_{\bar\theta} B_\mu)
- (d\bar\theta \wedge d \bar\theta) (\partial_{\bar\theta} \Phi)
- (d\theta \wedge d \bar\theta) (\partial_\theta \Phi +
\partial_{\bar\theta} \bar \Phi), \nonumber\\
\tilde A \wedge \tilde A &=& (dx^\mu \wedge dx^\nu) (B_\mu B_\nu)
+ (dx^\mu \wedge d\theta) ( [ B_\mu, \bar\Phi ] )
- (d\theta \wedge d \theta) (\bar \Phi \bar\Phi) \nonumber\\
&+& (dx^\mu \wedge d \bar\theta) ( [ B_\mu, \Phi ] )
- (d \bar\theta \wedge d \bar\theta) (\Phi \Phi)
- (d \theta \wedge d \bar\theta)(\{\Phi, \bar \Phi \}).
\end{array} \eqno(3.34)
$$
Ultimately, the super expansion in (2.46) (with group valued fields) can be
expressed in terms of the auxiliary fields in (3.33) and basic fields, as
$$
\begin{array}{lcl}
B_{\mu} (x, \theta, \bar \theta) &=& A_{\mu} (x) 
+ \theta\; \; D_{\mu} \bar C (x) + \bar \theta\; D_{\mu} C (x) 
+ i \;\theta \;\bar \theta \;(D_{\mu} B - i (D_\mu C \times \bar C)) (x), 
\nonumber\\
\Phi (x, \theta, \bar \theta) &=& C (x) 
+ i\; \theta  \bar B (x)
- \frac{\bar \theta}{2} \; (C \times C) (x) 
+ i\; \theta\; \bar \theta \;(\bar B \times  C) (x), \nonumber\\
\bar \Phi (x, \theta, \bar \theta) &=& \bar C (x) 
- \frac{\theta}{2} \; (\bar C \times \bar C) (x) 
+ i \; \bar \theta \;B (x) 
- i \;\theta \;\bar \theta \;(B \times \bar C) (x),
\end{array} \eqno(3.35)
$$
which can be recast in exactly the same form as (2.44) 
with $s_{(a)b} \rightarrow \tilde s_{(a)b}$ 
and $\tilde s_{(a)b}$ standing for the off-shell nilpotent
(anti-)BRST symmetries ((3.7) and (3.4))
for the non-Abelian gauge theories. {\it This form 
of super expansion provides the geometrical interpretation for the 
off-shell nilpotent (anti-)BRST 
charges $Q_{(a)b}$ as the translation generators along the Grassmannian 
direction $(\theta)\bar\theta$ of the supermanifold.} Similarly, the dual
horizontality condition $\tilde \delta \tilde A = \delta A$ (with
$ \delta = - * d *, \; \tilde \delta = - \star \tilde d \star$) leads to
the following relationships [39]
$$
\begin{array}{lcl}
R_\mu (x) &=& - \varepsilon_{\mu\nu} \partial^\nu \bar C (x), \quad
\bar R_\mu (x) = - \varepsilon_{\mu\nu} \partial^\nu C, \quad
B (x) =  \bar B (x) = 0, \nonumber\\
S_{\mu} (x) &=& + \varepsilon_{\mu\nu} \partial^\nu {\cal B} (x), \quad
{\cal B} (x) + \bar {\cal B} (x) = 0, \qquad  s (x) = \bar s(x) = 0.
\end{array} \eqno(3.36)
$$
In the above computation, we have used the following
$$
\begin{array}{lcl}
\tilde \delta \tilde A &=& 
\partial_\mu B^\mu + s^{\theta\theta} (\partial_{\theta} \Phi)
- \varepsilon^{\mu\theta} (\partial_\mu \Phi + \varepsilon_{\mu\nu}
\partial_{\theta} B^\nu) + s^{\bar\theta\bar\theta}
(\partial_{\bar\theta} \bar \Phi) \nonumber\\
&-& \varepsilon^{\mu \bar \theta} (\partial_\mu \bar \Phi 
+ \varepsilon_{\mu\nu} \partial_{\bar\theta} B^\nu)
+ s^{\theta\bar\theta} (\partial_\theta \bar \Phi 
+ \partial_{\bar\theta} \Phi).
\end{array} \eqno(3.37)
$$
Now the super expansion, after the application of the dual horizontality
condition, looks in terms of the (anti-)co-BRST transformations
$\tilde s_{(a)d}$ for non-Abelian gauge theory as 
$$
\begin{array}{lcl}
B_{\mu} (x, \theta, \bar \theta) &=& A_{\mu} (x) 
- \theta\; \varepsilon_{\mu\nu} \partial^\nu C (x) 
+ \bar \theta\; \varepsilon_{\mu\nu} \partial^\nu \bar C (x) 
+ i \;\theta \;\bar \theta \varepsilon_{\mu\nu} \partial^\nu {\cal B} (x), 
\nonumber\\
&\equiv& A_\mu (x) + \theta\; (\tilde s_{ad} A_\mu (x)) + \bar\theta\; 
(\tilde s_{d} A_\mu (x)) + \theta \bar\theta\; 
(\tilde s_{d} \tilde s_{ad} A_\mu (x)),\nonumber\\
\Phi (x, \theta, \bar \theta) &=& C (x) 
- i \;\bar \theta  \; {\cal B} (x) \equiv C (x) + \bar\theta \; 
(\tilde s_{d} C(x)), \nonumber\\
\bar \Phi (x, \theta, \bar \theta) &=& \bar C (x) 
+ i \; \theta \;{\cal B } (x) \equiv \bar C (x) + \theta\; 
(\tilde s_{ad} \bar C(x)),
\end{array} \eqno(3.38)
$$
which can be, finally, recast in exactly the same form as (2.45). This happens
here, unlike our earlier work [39], because of our choice of the 
signs in (2.46).
{\it It is very interesting to note that after the application of horizontality
condition for the derivation of the (anti-) BRST symmetries, the odd
superfield $\Phi$ and $\bar\Phi$ remain general 
({\bf not} (anti-)chiral) superfields. However, after
the application of dual horizontality condition for the derivation of the
(anti-)co-BRST symmetries, the odd super fields $(\bar \Phi)\Phi$ become
(anti-)chiral.} It is also straightforward to check that 

$$
\begin{array}{lcl}
\mbox{ Lim}_{\theta, \bar\theta \rightarrow 0}\;\;
{\displaystyle \frac{\partial}{\partial \bar \theta}}\; 
\tilde A_{M} (x,\theta, \bar\theta)
&=& - i \;\bigl [ \Psi (x), Q_{b} \bigr ]_{\pm} 
\equiv \tilde s_{b} \Psi (x), \nonumber\\
\mbox{ Lim}_{\theta, \bar\theta \rightarrow 0}\;\;
{\displaystyle \frac{\partial}{\partial \theta}}\; 
\tilde A_{M} (x,\theta, \bar\theta)
&=& - i \;\bigl [ \Psi (x), Q_{d} \bigr ]_{\pm} \equiv \tilde s_{d} \Psi (x).
\end{array} \eqno(3.39)
$$
which imply that
there are mappings between (anti-)BRST and (anti-)co-BRST transformations
of the local fields on the one hand and the translations of superfields
along Grassmannian directions of the $(2+2)$-dimensional supermanifold
on the other hand, as 
$$
\begin{array}{lcl}
\tilde s_{(d)b} \leftrightarrow \mbox{ Lim}_{\theta, \bar\theta \rightarrow 0}\;
{\displaystyle \frac{\partial}{\partial \bar \theta}}\; \qquad
\tilde s_{ab} \leftrightarrow  \mbox{ Lim}_{\theta, \bar\theta \rightarrow 0}\;
{\displaystyle \frac{\partial}{\partial \theta}}\; \qquad
\tilde s_{ad} \leftrightarrow  \mbox{ Lim}_{\theta, \bar\theta \rightarrow 0}\;
{\displaystyle \frac{\partial}{\partial \theta}}.
\end{array} \eqno(3.40)
$$
In spite the above similarity in the mapping, there is a clear-cut difference 
between $\tilde s_{b}$ and $\tilde s_{d}$ on the one hand and between 
$\tilde s_{ab}$ and $\tilde s_{ad}$
on the other hand as far as transformations on the (anti-)ghost
fields $(\bar C)C$ are concerned. To check the sanctity of (3.40), it is 
straightforward to note that in (3.10) we have the following equality
$$
\begin{array}{lcl}
\tilde s_{b} \Bigl (-i \bar C \cdot [ \partial_\mu A^\mu + \frac{1}{2}
\xi B ] \Bigr ) = \mbox{Lim}_{\theta,\bar\theta \rightarrow 0}\;
{\displaystyle \frac{\partial}{\partial \bar\theta}} \Bigl [
- i \bar\Phi \cdot \partial_\mu B^\mu + \frac{1}{2} \xi \bar \Phi
\cdot {\displaystyle \frac{\partial \bar\Phi}{\partial \bar\theta} }
\Bigr ]|_{\mbox{(anti-)BRST}},
\end{array} \eqno(3.41)
$$
where the subscript stands for the expansion of the superfields in (3.35). 
Similarly, it can be seen that the BRST exact quantity in (3.11) is
$$
\begin{array}{lcl}
\tilde s_{b} \tilde s_{ab} \Bigl ( \frac{i}{2} A_\mu \cdot A^\mu
- \xi \bar C \cdot C  \Bigr ) = {\displaystyle 
\frac{\partial}{\partial\bar\theta}
\frac{\partial}{\partial\theta} } \Bigl (\frac{i}{2} B_\mu \cdot B^\mu 
- \xi \bar\Phi \cdot \Phi \Bigr )|_{\mbox{(anti-)BRST}}, 
\end{array}\eqno(3.42)
$$
and BRST co-exact- and anti-co-exact quantities of (3.11) are
$$
\begin{array}{lcl}
\tilde s_{d} \Bigl (\;i C \cdot [\;E - \frac{1}{2} {\cal B}\; ]\;\Bigr ) &=&
\mbox{Lim}_{\theta,\bar\theta \rightarrow 0} 
{\displaystyle \frac{\partial}{\partial\bar\theta}} 
\Bigl [ - i \Phi \cdot
\varepsilon^{\mu\nu} (\partial_\mu B_\nu + \frac{1}{2} B_\mu \times B_\nu )
\nonumber\\
&+& {\displaystyle
 \frac{1}{2} \Phi \cdot \frac{\partial \Phi}{\partial\bar\theta}} 
\Bigr ]|_{\mbox{(anti-)co-BRST}} \nonumber\\
&\equiv& \mbox{Lim}_{\theta,\bar\theta \rightarrow 0} \Bigl [ - i \Phi \cdot
\varepsilon^{\mu\nu} (\partial_\mu B_\nu + \frac{1}{2} B_\mu \times B_\nu )
\nonumber\\
&-& {\displaystyle
 \frac{1}{2} \Phi \cdot \frac{\partial \bar\Phi}{\partial \theta}} 
\bigr ]|_{\mbox{(anti-)co-BRST}},
\end{array}\eqno(3.43)
$$
$$
\begin{array}{lcl}
\tilde s_{ad} \Bigl (\;- i \bar C \cdot [\;E 
- \frac{1}{2} {\cal B}\; ]\; \Bigr ) &=&
\mbox{Lim}_{\theta,\bar\theta \rightarrow 0} 
{\displaystyle \frac{\partial}{\partial\theta}} 
\Bigl [ + i \bar \Phi \cdot
\varepsilon^{\mu\nu} (\partial_\mu B_\nu + \frac{1}{2} B_\mu \times B_\nu )
\nonumber\\
&-& {\displaystyle
 \frac{1}{2} \bar \Phi \cdot \frac{\partial \Phi}{\partial\bar\theta}} 
\Bigr ]|_{\mbox{(anti-)co-BRST}} \nonumber\\
&\equiv& \mbox{Lim}_{\theta,\bar\theta \rightarrow 0} 
{\displaystyle \frac{\partial}{\partial\theta}} 
\Bigl [ + i \bar\Phi \cdot
\varepsilon^{\mu\nu} (\partial_\mu B_\nu + \frac{1}{2} B_\mu \times B_\nu )
\nonumber\\
&+& {\displaystyle
 \frac{1}{2} \Phi \cdot \frac{\partial \bar\Phi}{\partial \theta}} 
\Bigr ]|_{\mbox{(anti-)co-BRST}},
\end{array} \eqno(3.44)
$$
where the subscripts stand for the expansions in (3.38).\\

\noindent
{\bf 4 Topological aspects: superfield formulation}\\

\noindent
We deal here with the topological features of the
free 2D Abelian- and self-interacting non-Abelian  gauge theories in the
framework of superfield formalism.\\

\noindent
{\bf 4.1 Topological features of free 2D Abelian theory: superfield approach}\\

\noindent
We explore here the possibilities of expressing some of the topological
aspects of the free 2D Abelian gauge theory in terms of (anti-)chiral 
superfields as well as general superfields which have been responsible for
our derivation of the on-shell and off-shell nilpotent symmetries. We 
also provide here geometrical interpretation for some of the topologically
interesting quantities on the supermanifold. Let us first concentrate on
the appearance of the Lagrangian density (2.1) (in its new form (2.7)) which 
respects the on-shell nilpotent (anti-)BRST and (anti-)co-BRST symmetries.
With the help of the (anti-)chiral superfields, it can be checked that (2.7)
can be expressed as

$$
\begin{array}{lcl}
{\cal L}_{b} + \partial_\mu X^\mu 
= - {\displaystyle \frac{i}{2}\; \frac{\partial}{\partial \bar \theta}}
\Bigl [\; \{ (\varepsilon^{\mu\nu} \partial_\mu B_\nu) \Phi \}|_{\mbox{co-BRST}}
+ \{ (\partial_\mu B^\mu) \bar \Phi \}|_{\mbox{BRST}}\; \Bigr ],
\end{array} \eqno(4.1)
$$
$$
\begin{array}{lcl}
{\cal L}_{b} + \partial_\mu X^\mu 
= + {\displaystyle \frac{i}{2}\; \frac{\partial}{\partial \theta}}
\Bigl [\; \{ (\varepsilon^{\mu\nu} \partial_\mu B_\nu) 
\bar \Phi \}|_{\mbox{anti-co-BRST}}
+ \{ (\partial_\mu B^\mu) \Phi \}|_{\mbox{anti-BRST}}\; \Bigr ],
\end{array} \eqno(4.2)
$$
where the subscripts stand for the substitution of the super expansions in
(2.19), (2.26), (2.33) and (2.40) and $X^\mu = \frac{i}{2} (\partial^\mu \bar
C C + \bar C \partial^\mu C) \equiv \frac{i}{2} \partial^\mu (\bar C C)$.
There is a symmetry that relates (4.1) and (4.2). In fact, these equations
 exchange with
each-other under: $\theta \leftrightarrow - \bar\theta, \Phi \leftrightarrow
\bar \Phi$ while the subscripts `BRST' and `anti-BRST' as well as `co-BRST'
and `anti-co-BRST' are exchanged with each-other.
Mathematically, the above equation (4.1) implies that the Lagrangian density
(2.1) is the $\bar\theta$-component of the composite {\it chiral} superfields
$(\varepsilon^{\mu\nu} \partial_\mu B_\nu) \Phi$ and $(\partial_\mu B^\mu) 
\bar \Phi$ with the substitution of expansions derived after the application
of (dual) horizontality conditions. The same (2.1) is also equivalent to
the $\theta$-component of the similar kind of {\it anti-chiral} superfields 
(with the
exchange of the fermionic superfields $\Phi \leftrightarrow \bar \Phi$)
and for these superfields, the expansions are the ones derived after the
application of (dual) horizontality conditions. In the language of the
mapping: $s_{(d)b} \leftrightarrow \partial / \partial \bar \theta$, 
it is clear that
geometrically the Lagrangian density ${\cal L}_{b}$ is equivalent to the
sum of two terms that correspond to the translations of the composite 
{\it chiral} superfields $(\varepsilon^{\mu\nu} \partial_\mu B_\nu) \Phi$ and
$(\partial_\mu B^\mu)\bar\Phi$ along the $\bar\theta$-direction of the
supermanifold. This translation is generated by the on-shell nilpotent
(co-)BRST charges. Similarly, in view of the mapping $s_{ab} \leftrightarrow
\partial / \partial \theta, s_{ad}  \leftrightarrow 
\partial / \partial \theta$, it is straightforward
to check that (2.1) is also equivalent to the translation of the
{\it anti-chiral} composite superfields 
$(\varepsilon^{\mu\nu} \partial_\mu B_\nu)
\bar\Phi$ and $(\partial_\mu B^\mu) \Phi$ along the $\theta$-direction
of the supermanifold. The corresponding translations are generated by
the on-shell nilpotent anti-BRST and anti-co-BRST charges. There is another
way to look geometrically at the Lagrangian density (2.1) in the light of
super expansions in (2.43). In fact, in addition to (4.2) and (4.1), there
is a completely {\it new} way  to express (2.1)
$$
\begin{array}{lcl}
{\cal L}_{b} + \partial_\mu X^\mu 
= + {\displaystyle \frac{i}{4}\; \frac{\partial}{\partial {\bar\theta}}
\;\frac{\partial}{\partial {\theta}}}\;
\Bigl [\; \bigl (B_\mu B^\mu \bigr)|_{\mbox{(anti-)co-BRST}}
+ \bigl ( B_\mu B^\mu \bigr )|_{\mbox{(anti-)BRST}}\; \Bigr ],
\end{array} \eqno(4.3)
$$
because the super expansion for the bosonic field $B_\mu$, vis-a-vis
(2.43), are
$$
\begin{array}{lcl}
&&B_\mu (x,\theta,\bar\theta)|_{\mbox{(anti-)co-BRST}} 
= A_\mu (x) - \theta \varepsilon_{\mu\nu} \partial^\nu  C (x)
- \bar\theta \varepsilon_{\mu\nu} \partial^\nu \bar C (x) 
+ i \theta \bar\theta\; \varepsilon_{\mu\nu} \partial^\nu E (x), \nonumber\\
&&B_\mu (x,\theta,\bar\theta)|_{\mbox {(anti-)BRST}}
 = A_\mu (x) + \theta\; \partial_\mu \bar C (x)
+ \bar\theta\; \partial_\mu C (x) - \frac{i}{\xi} \;\theta\;\bar\theta\; 
\partial_\mu (\partial_\rho A^\rho) (x).
\end{array} \eqno(4.4)
$$
In the language of the (anti-)BRST $s_{(a)b}$ and (anti-)co-BRST $s_{(a)d}$
transformations on the basic fields of the Lagrangian density (2.1), the
above superfield formulation implies
$$
\begin{array}{lcl}
{\cal L}_{b} + \partial_\mu X^\mu 
= + \frac{i}{4}\; \bigl [\; s_{b} s_{ab}\;(A_\mu A^\mu) + 
s_{d} s_{ad} \;(A_\mu A^\mu)\; \bigr ],
\end{array} \eqno(4.5)
$$
where $X^\mu = \frac{1}{2}\;[ \frac{1}{\xi}\; A^\mu (\partial_\rho A^\rho) 
+ \varepsilon^{\mu\nu} A_\nu E ]$
for both (4.3) and (4.5). Thus, it can be emphasized, at this stage,
that {\it the superfield formulation provides a new insight into the topological
nature of the Lagrangian density (2.1) as it can be expressed in a way
that was not known hitherto in our previous works [29-35]}. The equations (4.3)
and (4.5) are endowed with symmetries that are inter-related. The invariance
of (4.3) under $\theta \leftrightarrow - \bar\theta$ is reflected in the 
invariance of (4.5) under: $s_{b} \leftrightarrow - s_{ab}, s_{d} 
\leftrightarrow - s_{ad}$. Mathematically,
the Lagrangian density (2.1) can be thought of as the 
$\theta\bar\theta$-component of the scalar superfield 
($B_\mu B^\mu$) {\it constructed by the 
bosonic superfields} $B_\mu$ for which the expansions in (4.4) have to be
plugged in. In the language of geometry on the supermanifold, the Lagrangian
density (2.1) is equal to the sum of two terms that correspond to a couple
of successive translations of the Lorentz superscalar $(B_\mu B^\mu)$
along the $\theta$- and $\bar\theta$-directions of
the supermanifold. These translations of the superscalar
are generated by the (anti-)BRST and (anti-)co-BRST charges which ultimately
correspond to the sum of on-shell nilpotent
(anti-)BRST and (anti-)co-BRST transformations on the
Lorentz scalar $(A_\mu A^\mu)$ as is evident from (4.5).

We shall concentrate now on the topological invariants 
(in particular the zero-forms) of the theory in the
language of the superfield formulation. It can be checked that, for the choice
of the (anti-)chiral superfields which lead to the derivation of the on-shell
nilpotent (co-)BRST symmetries in (2.26) and (2.19) (and anti-BRST and 
anti-co-BRST symmetries in (2.33) and (2.40)), we have the following
$$
\begin{array}{lcl}
&&(\Phi \bar\Phi)|_{\mbox{BRST}} = C \bar C + \frac{i}{\xi}\; \bar \theta\; C 
(\partial_\mu A^\mu), \quad
(\Phi \bar\Phi)|_{\mbox{anti-BRST}} = C \bar C + \frac{i}{\xi}\;\theta \; 
\bar C (\partial_\mu A^\mu), \nonumber\\
&&(\Phi \bar\Phi)|_{\mbox{co-BRST}} =  C \bar C 
- i \; \bar \theta\; E \bar C,  \quad
(\Phi \bar\Phi)|_{\mbox{anti-co-BRST}} = C \bar C 
- i \;\theta \;  E C,
\end{array} \eqno(4.6)
$$
where the subscripts have the same meaning as explained earlier.
From the above equations, it can be seen that we obtain the following 
zero-forms quoted in (2.12)
$$
\begin{array}{lcl}
{\displaystyle i \frac{\partial}{\partial\bar\theta}\; }
(\Phi\bar\Phi)|_{\mbox{BRST}} &=&- \frac{1}{\xi}\;
(\partial_\mu A^\mu) C \equiv V_{0},\quad 
{\displaystyle i \frac{\partial}{\partial\bar\theta}\; }
(\Phi\bar\Phi)|_{\mbox{co-BRST}} =
E \bar C \equiv W_{0},\nonumber\\
{\displaystyle i \frac{\partial}{\partial\theta}\;}
 (\Phi\bar\Phi)|_{\mbox{anti-BRST}} 
&=& - \frac{1}{\xi}\;
(\partial_\mu A^\mu) \bar C \equiv \bar V_{0}, \quad
{\displaystyle i \frac{\partial}{\partial\theta}\; }
(\Phi\bar\Phi)|_{\mbox{anti-co-BRST}} 
=  E C \equiv \bar W_{0}. 
\end{array} \eqno(4.7)
$$
In the language of transformations on the basic fields of (2.1), we have
$$
\begin{array}{lcl}
V_{0} = s_{b} (i C \bar C), \quad \bar V_{0} = s_{ab} (i C \bar C), \quad
W_{0} = s_{d} (i C \bar C), \quad \bar W_{0} = s_{ad} (i C \bar C),
\end{array} \eqno(4.8)
$$
which clearly and explicitly show that $(\bar V_{0})V_{0}$ are the (anti-)BRST 
invariant and $(\bar W_{0}) W_{0}$ are the (anti-)co-BRST invariant
quantities due to the on-shell nilpotency ($s_{(a)b}^2 = 0, s_{(a)d}^2 = 0$) of
these charges. In the language of the superspace variables, the above
nilpotency is encoded in the nilpotency of the derivatives
($(\partial/ \partial\theta)^2 = 0, (\partial/ \partial \bar\theta)^2 = 0$) 
w.r.t. the Grassmannian
variables $\theta$ and $\bar\theta$. The above zero-forms, that have been
computed by taking into account separately (anti-)chiral superfields, can
be obtained {\it together} by considering the super expansion (2.43) where
(anti-)chiral superfields are merged together consistently. 
This expansion, after the
application of (dual) horizontality conditions, yields the followings
$$
\begin{array}{lcl}
(\Phi \bar \Phi)|_{\mbox{(anti-)co-BRST}}
 &=& C \bar C - i \theta C E - i \bar\theta E \bar C
- \theta \bar\theta E^2, \nonumber\\
(\Phi \bar \Phi)|_{\mbox{(anti-)BRST}}
 &=& C \bar C + \frac{i}{\xi} \theta (\partial_\mu A^\mu) \bar C
+ \frac{i}{\xi}\; \bar\theta (\partial_\mu A^\mu) C 
+ \frac{1}{\xi^2}\; \theta\bar\theta \;(\partial_\mu A^\mu)^2.
\end{array} \eqno(4.9)
$$
From the above expansions, we obtain the invariant quantities of (2.12), as
$$
\begin{array}{lcl}
\mbox{Lim}_{\theta,\bar\theta \rightarrow 0}\;\;
{\displaystyle i \frac{\partial}{\partial\bar\theta}\; }
(\Phi\bar\Phi)|_{\mbox{(anti-)BRST}} &=&- \frac{1}{\xi}\;
(\partial_\mu A^\mu) C \equiv V_{0},\nonumber\\
\mbox{Lim}_{\theta,\bar\theta \rightarrow 0}\;\;
{\displaystyle i \frac{\partial}{\partial \theta}\; }
(\Phi\bar\Phi)|_{\mbox{(anti-)BRST}} 
&=& - \frac{1}{\xi}\;
(\partial_\mu A^\mu) \bar C \equiv \bar V_{0}, \nonumber\\
\mbox{Lim}_{\theta,\bar\theta \rightarrow 0}\;\;
{\displaystyle i \frac{\partial}{\partial\bar\theta}\; }
(\Phi\bar\Phi)|_{\mbox{(anti-)co-BRST}} &=&
E \bar C \equiv W_{0}, \nonumber\\
\mbox{Lim}_{\theta,\bar\theta \rightarrow 0}\;\;
{\displaystyle i \frac{\partial}{\partial \theta}\; }
(\Phi\bar\Phi)|_{\mbox{(anti-)co-BRST}} 
&=&  E C \equiv \bar W_{0}. 
\end{array} \eqno(4.10)
$$
It is clear from equations (4.7) and (4.10) that, mathematically, the 
zero-forms of the topological invariants w.r.t. on-shell nilpotent
(anti-)BRST and (anti-)co-BRST charges are $(\theta)\bar\theta$ components
of a bosonic superfield, constituted {\it by merging a couple of fermionic
superfields $\Phi$ and $\bar\Phi$ in the composite form $\Phi\bar\Phi$}. In the
language of geometry on the supermanifold, the above zero-forms are nothing but
the translations of the composite superfield $\Phi\bar\Phi$ along the
$(\theta)\bar\theta$-directions of the supermanifold by the on-shell nilpotent
(anti-)BRST and (anti-)co-BRST charges. As far as 
the topological invariants w.r.t.
off-shell nilpotent (anti-)BRST and (anti-)co-BRST charges are concerned, we
notice from (2.48) and (2.50) 
$$
\begin{array}{lcl}
(\Phi \bar \Phi)|_{\mbox{(anti-)co-BRST}}
 &=& C \bar C - i \theta C {\cal B} - i \bar\theta {\cal B} \bar C
- \theta \bar\theta {\cal B}^2, \nonumber\\
(\Phi \bar \Phi)|_{\mbox{(anti-)BRST}}
 &=& C \bar C - i \theta B \bar C
- i \bar\theta  B C 
+ \theta\bar\theta \;B^2.
\end{array} \eqno(4.11)
$$
It is straightforward to check that the invariant quantities of (2.12) are
$$
\begin{array}{lcl}
\mbox{Lim}_{\theta,\bar\theta \rightarrow 0}\;\;
{\displaystyle i \frac{\partial}{\partial\bar\theta}\; }
(\Phi\bar\Phi)|_{\mbox{(anti-)BRST}} &=& B C \equiv V_{0}, \nonumber\\
\mbox{Lim}_{\theta,\bar\theta \rightarrow 0}\;\;
{\displaystyle i \frac{\partial}{\partial \theta}\; }
(\Phi\bar\Phi)|_{\mbox{(anti-)BRST}} 
&=& B \bar C \equiv \bar V_{0}, \nonumber\\
\mbox{Lim}_{\theta,\bar\theta \rightarrow 0}\;\;
{\displaystyle i \frac{\partial}{\partial\bar\theta}\; }
(\Phi\bar\Phi)|_{\mbox{(anti-)co-BRST}} &=&
{\cal B} \bar C \equiv W_{0}, \nonumber\\
\mbox{Lim}_{\theta,\bar\theta \rightarrow 0}\;\;
{\displaystyle i \frac{\partial}{\partial \theta}\; }
(\Phi\bar\Phi)|_{\mbox{(anti-)co-BRST}} 
&=& {\cal B} C \equiv \bar W_{0}. 
\end{array} \eqno(4.12)
$$
The mathematical as well as the geometrical interpretations for the above
zero-forms can be given in a similar manner as has been given for equations
(4.7) and (4.8). Only the difference here is that the generators
for the internal transformations ($\tilde s_{(a)b}, \tilde s_{(a)d}$) 
on the basic
fields $\Psi$ of the Lagrangian density (2.3) and corresponding translation
generators ($\frac{\partial}{\partial \theta}, 
\frac{\partial}{\partial \bar \theta}$) on the supermanifold are the 
{\it off-shell} nilpotent (anti-)BRST and (anti-)co-BRST charges. It is
very interesting to note that (i) the zero-form(s) of the topological 
invariants and the Lagrangian density (cf. (4.3)) correspond to the
Grassmannian translations on the local but composite {\it even} superfields,
and (ii) the composite {\it even} superfields $(\Phi\bar\Phi)$ in the
former and $(B^\mu B_\mu)$ (i.e., a Lorentz scalar)
in the latter are constituted by the {\it odd}
scalar superfields and the {\it even} vector superfields of the theory.

We shall focus on now the symmetric energy momentum tensor 
$(T_{\mu\nu}^{(s)})$ for the above theory. In terms
of the chiral superfield expansions in (2.19) and (2.26), we have the
following form, modulo some total derivatives, for $T_{\mu\nu}^{(s)}$ in (2.9)
$$
\begin{array}{lcl}
T_{\mu\nu}^{(s)} &=& {\displaystyle \frac{i}{2}\; \frac{\partial}
{\partial \bar\theta}}
\Bigl [\; P_{\mu\nu}^{(s)}|_{\mbox{BRST}} \; + \; 
Q_{\mu\nu}^{(s)}|_{\mbox{co-BRST}} \; \Bigr ], \nonumber\\
 P_{\mu\nu}^{(s)} &=& 
\partial_{\mu} \bar \Phi B_\nu
+ \partial_\nu \bar \Phi B_\mu + \eta_{\mu\nu} (\partial_\rho B^\rho) 
\bar \Phi,\nonumber\\
Q_{\mu\nu}^{(s)} &=&
\partial_{\mu} \Phi \varepsilon_{\nu\rho} B^\rho 
+ \partial_\nu \Phi \varepsilon_{\mu\rho} B^\rho 
+ \eta_{\mu\nu} \varepsilon^{\rho\sigma} \partial_\rho B_\sigma \Phi.
\end{array} \eqno(4.13)
$$
The above energy-momentum tensor,
modulo some total derivatives, can also be expressed in terms of the
anti-chiral super expansions in (2.33) and (2.40), as 
$$
\begin{array}{lcl}
T_{\mu\nu}^{(s)} &=& - {\displaystyle \frac{i}{2}\; \frac{\partial}
{\partial \theta}}
\Bigl [\; R_{\mu\nu}^{(s)}|_{\mbox{anti-BRST}} \; + \; 
S_{\mu\nu}^{(s)}|_{\mbox{anti-co-BRST}} \; \Bigr ], \nonumber\\
 R_{\mu\nu}^{(s)} &=& 
\partial_{\mu} \Phi B_\nu
+ \partial_\nu  \Phi B_\mu + \eta_{\mu\nu} (\partial_\rho B^\rho) 
\Phi,\nonumber\\
S_{\mu\nu}^{(s)} &=&
\partial_{\mu} \bar \Phi \varepsilon_{\nu\rho} B^\rho 
+ \partial_\nu \bar \Phi \varepsilon_{\mu\rho} B^\rho 
+ \eta_{\mu\nu} \varepsilon^{\rho\sigma} \partial_\rho B_\sigma \bar \Phi.
\end{array} \eqno(4.14)
$$
Mathematically, the above equations (4.13) and (4.14) imply that, modulo
some total derivatives, $T_{\mu\nu}^{(s)}$ corresponds to the $\theta (\bar
\theta)$ components of the sum of (anti-)chiral superfields $(R)P$ and
$(S)Q$ respectively. Geometrically, the symmetric energy momentum tensor
corresponds to the translations of the sum of (anti-)chiral superfields $(R)P$ 
and $(S)Q$, along the $(\theta)\bar\theta$-directions of the 
$(2+2)$-dimensional supermanifold. These translations are generated by the
on-shell nilpotent (anti-)BRST and (anti-)co-BRST charges.
In view of the expansions in (2.43) where the (anti-)chiral super expansions
have been merged together in a consistent way, it can be checked that, besides
expressions in (4.13) and (4.14), {\it there is a novel way to express }
the symmetric energy momentum tensor $T_{\mu\nu}^{(s)}$ of (2.9), as
$$
\begin{array}{lcl}
&&T_{\mu\nu}^{(s)} = {\displaystyle \frac{i}{2}\;\frac{\partial}
{\partial\bar\theta}\; \frac{\partial}{\partial\theta}}\;
\bigl [\; (B_\mu B^\mu) (x,\theta,\bar\theta) 
- \frac{1}{2} \eta_{\mu\nu} (B_\rho B^\rho) (x,\theta,\bar\theta)
\;\bigr ]|_{\mbox{(anti-)BRST}} \nonumber\\
&-& {\displaystyle \frac{i}{2}\;\frac{\partial}
{\partial\bar\theta}\; \frac{\partial}{\partial\theta}}\;
\bigl [\; \varepsilon_{\mu\rho} \varepsilon_{\nu\sigma}
(B^\rho B^\sigma) (x,\theta,\bar\theta) 
+ \frac{1}{2} \eta_{\mu\nu} (B_\rho B^\rho) (x,\theta,\bar\theta)
\;\bigr ]|_{\mbox{(anti-)co-BRST}},
\end{array} \eqno(4.15)
$$
where the subscripts correspond to the modified expansion of (2.43) after
the application of the (dual) horizontality conditions (see, e.g.,
Ref. [40] for more details). As far as the transformations on the basic
fields of the Lagrangian density (2.1) are concerned, the above symmetric
energy-momentum (4.15), modulo some total derivatives, can be expressed as
$$
\begin{array}{lcl}
T_{\mu\nu}^{(s)} = \frac{i}{2}\; s_{b}s_{ab} \bigl (A_\mu A^\mu - \frac{1}{2}
\eta_{\mu\nu} A_\rho A^\rho \bigr ) - \frac{i}{2}\; s_{d}s_{ad}
\bigl ( \varepsilon_{\mu\rho} \varepsilon_{\nu\sigma} A^\rho A^\sigma
+ \frac{1}{2} \eta_{\mu\nu} A_\rho A^\rho \bigr ),
\end{array} \eqno(4.16)
$$
which encodes the topological nature of the theory in a grand and illuminating
manner in view of the mapping $s_{(a)b} \leftrightarrow Q_{(a)b}$ and
$s_{(a)d} \leftrightarrow Q_{(a)d}$. In other words, the above expression for
$T_{\mu\nu}^{(s)}$ shows that it is 
equal to the sum of two (anti-)commutators. It is
worth pointing out that in all our previous works [29-35], we were unable to
express the energy momentum tensor in the form presented in (4.16). Thus,
the study of the theory in the superfield formulation does lead to 
the discovery of
some new symmetries of the $T_{\mu\nu}^{(s)}$. As explained for the
equations (4.13) and (4.14), we can provide the
mathematical as well as geometrical
interpretation for the equation (4.15). Mathematically, the 
$\theta\bar\theta$-component of the superfields 
$B_\mu B_\nu - \frac{1}{2} \eta_{\mu\nu} B_\rho B^\rho$
and $\varepsilon_{\mu\rho} \varepsilon_{\nu\rho} B^\rho B^\sigma
+ \frac{1}{2} \eta_{\mu\nu} B_\rho B^\rho$ correspond to $T_{\mu\nu}^{(s)}$
of the theory when we substitute for the bosonic superfield $B_\mu$, 
the super expansions of (4.4) (that are
obtained after the imposition of the (dual) horizontality
conditions). In the language of the geometry on the supermanifold, the equation
(4.15) implies that $T_{\mu\nu}^{(s)}$ is equivalent to a couple
of successive translations of the superfields
$B_\mu B_\nu - \frac{1}{2} \eta_{\mu\nu} B_\rho B^\rho$ and
$\varepsilon_{\mu\rho} \varepsilon_{\nu\rho} B^\rho B^\sigma
+ \frac{1}{2} \eta_{\mu\nu} B_\rho B^\rho$ 
along $\theta$ and $\bar\theta$-directions of the
supermanifold. These translations are generated by the 
on-shell nilpotent (anti-)BRST and 
(anti-)co-BRST charges.\\

\noindent
{\bf 4.2 Topological features of non-Abelian theory: superfield formalism}\\

\noindent
We shall discuss here some of the key topological features of the 2D
self-interacting non-Abelian gauge theory. To start with, it is evident
from (3.9) that the Lagrangian density (3.1) turns out to be the sum of
two anti-commutators with the on-shell nilpotent (co-) BRST charges
$Q_{(d)b}$. This fact can be captured in terms of the chiral superfield
expansions in (3.18) and (3.22) that have been obtained after the
imposition of the (dual) horizontality conditions. In fact, in terms
of the composite chiral superfields, the Lagrangian density
(3.9) can be expressed,
modulo some total derivative $\partial_\mu X^\mu$,  as
$$
\begin{array}{lcl}
{\cal L}_{b}^{(N)}  
= - {\displaystyle \frac{i}{2} \frac{\partial}{\partial\bar\theta}}\;
\Bigl (\;[(\partial_\mu B^\mu) \cdot \bar \Phi]|_{\mbox{BRST}} 
+ [\varepsilon^{\mu\nu} (\partial_\mu B_\nu 
+ \frac{1}{2} \; B_\mu \times B_\nu) \cdot \Phi]|_{\mbox{co-BRST}} \Bigr ),
\end{array} \eqno(4.18)
$$
where $X^\mu = \frac{i}{2} (\bar C \cdot D^\mu C + \partial^\mu \bar C
\cdot C)$ and subscripts correspond to the insertions of super expansions
in (3.18) and (3.22) respectively. Mathematically, the Lagrangian density
(3.1) (or its analogue (3.9)) is nothing but the $\bar\theta$ component
of the local but composite chiral superfields $(\partial_\mu B^\mu) \cdot 
\bar \Phi$ and $\varepsilon^{\mu\nu} (\partial_\mu B_\nu + \frac{1}{2}
B_\mu \times B_\nu) \cdot \Phi$ where the insertions from 
(3.18) and (3.22) are made.
Recalling our result of the analogy $s_{(d)b} \leftrightarrow 
\partial / \partial \bar \theta$, it is very clear that (4.18) 
is nothing but (3.9)
where the basic fields have been replaced by the corresponding superfields.
In the language of geometry on the supermanifold, it can be noticed that
the translation of the local (but composite) chiral superfields 
$(\partial_\mu B^\mu) \cdot \bar \Phi$ and $\varepsilon^{\mu\nu} 
(\partial_\mu B_\nu + \frac{1}{2} B_\mu \times B_\nu) \cdot \Phi$ 
along the $\bar\theta$-direction of the supermanifold corresponds to 
the sum of on-shell
nilpotent transformations $s_{b}$ and $s_{d}$ on the local (but composite)
fields $- \frac{1}{2} (\partial_\rho A^\rho) \cdot \bar C$ and 
$\frac{1}{2} E \cdot C$, respectively. This observation implies the 
topological nature of the theory. In fact, in the language of chiral 
superfields, the Lagrangian density (4.18) of the
2D self-interacting non-Abelian gauge theory is nothing but the total
Grassmannian derivative w.r.t. $\bar\theta$ on the composite chiral
superfields (cf. (4.18)). Here the physical states 
(as well as the vacuum) of the theory are supposed to be annihilated by the 
conserved and nilpotent (co-)BRST charges because they are the harmonic state
of any arbitrary Hodge decomposed state of the quantum Hilbert space.

Let us now concentrate on the topological invariants of the theory. In
particular, we shall provide the geometrical interpretation 
(in the language of translations on the supermanifold) for the
on-shell ($D_\mu \partial^\mu \bar C = \partial_\mu D^\mu C = 0$)
(co-)BRST invariant quantities connected with the zero-forms of the
topological invariants, defined on the 2D manifold. The higher degree
forms (i.e. one- and two-forms) can be computed from the zero-form
by exploiting the recursion relations [35]. Towards this goal in mind, let us
note the following
$$
\begin{array}{lcl}
(\Phi \cdot \bar \Phi)|_{\mbox{BRST}}
&=& C \cdot \bar C + \frac{i}{\xi} \bar\theta C \cdot (\partial_\mu A^\mu)
- \frac{1}{2}\;\bar\theta\; \bar C \cdot (C \times C), \nonumber\\
(\Phi \cdot \bar \Phi)|_{\mbox{co-BRST}}
&=& C \cdot \bar C - i \bar\theta\; E \cdot \bar C,
\end{array} \eqno(4.19)
$$
where the subscripts stand for the insertions of the chiral superfield
expansions in (3.18) and (3.22) that have been obtained after the
imposition of the (dual) horizontality conditions. It is straightforward to
check that the invariant quantities in (3.14) are
$$
\begin{array}{lcl}
{\displaystyle i \frac{\partial}{\partial\bar\theta}\; }
(\Phi \cdot \bar\Phi)|_{\mbox{BRST}} &=& - \frac{1}{\xi}
C \cdot (\partial_\mu A^\mu)
- \frac{i}{2}\; \bar C \cdot (C \times C) \equiv V_{0}, \nonumber\\ 
{\displaystyle i \frac{\partial}{\partial\bar\theta}\; }
(\Phi \cdot \bar\Phi)|_{\mbox{co-BRST}} &=&
\;\;E \cdot \bar C \;\;\equiv \;\;W_{0}, 
\end{array} \eqno(4.20)
$$
which remain on-shell ($D_\mu \partial^\mu \bar C = \partial_\mu D^\mu C = 0$)
BRST and co-BRST invariant, respectively. Mathematically,  
the $\theta$- and $\bar\theta$-
components of the chiral superfields $(i \Phi \cdot \bar \Phi)$ lead to
the derivation of zero-forms which are the on-shell (co-)BRST invariant
quantities $(W_{0})V_{0}$. Geometrically, the zero-forms $(W_{0})V_{0}$
w.r.t. the on-shell  nilpotent (co-)BRST charges are nothing but the
translations of the chiral superfields $(i \Phi \cdot \bar\Phi)$
along the $(\theta)\bar\theta$-directions of the supermanifold. For the 
computation of
the off-shell invariant (anti-)BRST and (anti-) co-BRST quantities, we shall
concentrate on the expansion (2.46) (where all the fields are group valued
e.g. $A_\mu = A_\mu^a T^a$ etc.) and compute the analogues of (4.20) from the
general expansions in (3.35) and (3.38). Thus, the invariant
quantities in (3.15) are
$$
\begin{array}{lcl}
{\displaystyle i \frac{\partial}{\partial\bar\theta}\; }
(\Phi \cdot \bar\Phi)|_{\mbox{(anti-)BRST}} &=& B \cdot C - \frac{i}{2} 
\bar C \cdot (C \times C) \equiv V_{0} 
\nonumber\\
{\displaystyle i \frac{\partial}{\partial \theta}\; }
(\Phi \cdot \bar\Phi)|_{\mbox{(anti-)BRST}} 
&=& - \bar B \cdot \bar C + \frac{i}{2} C \cdot (\bar C \times \bar C)
\equiv \bar V_{0}, 
\nonumber\\
{\displaystyle i \frac{\partial}{\partial\bar\theta}\; }
(\Phi \cdot \bar\Phi)|_{\mbox{(anti-)co-BRST}} &=&
{\cal B} \cdot \bar C \equiv W_{0}, \nonumber\\
{\displaystyle i \frac{\partial}{\partial \theta}\; }
(\Phi \cdot \bar\Phi)|_{\mbox{(anti-)co-BRST}} &=& {\cal B} \cdot C
\equiv \bar W_{0}.
\end{array}\eqno(4.21)
$$
The mathematical as well as geometrical interpretations of the 
above zero-forms will go along the same lines as that for (4.20).

Now we dwell a bit on the form of the symmetric energy momentum tensor. In
terms of the on-shell nilpotent (co-)BRST symmetries and corresponding chiral
super expansion of the superfields in (3.18) and (3.22), it can be seen that
$T_{\alpha\beta}^{(N)}$ of equation (3.12) can be written, modulo
some total derivatives $X_{\alpha\beta}^{(N)}$, as
$$
\begin{array}{lcl}
T_{\alpha\beta}^{(N)} + X_{\alpha\beta}^{(N)}
&=& {\displaystyle \frac{i}{2} \frac{\partial}{\partial\bar\theta}}
\Bigl [\; Y_{\alpha\beta}^{(N)}|_{\mbox{BRST}} \; 
+ \; Z_{\alpha\beta}^{(N)}|_{\mbox{co-BRST}} \Bigr ],
\nonumber\\
Y_{\alpha\beta}^{(N)} &=&
\partial_\alpha \bar \Phi \cdot B_\beta + \partial_\beta \bar \Phi \cdot
B_\alpha + \eta_{\alpha\beta} (\partial_\rho A^\rho) \cdot \bar \Phi,
\nonumber\\
Z_{\alpha\beta}^{(N)} &=&
\partial_\alpha \Phi \cdot \varepsilon_{\beta\rho}
B^\rho + \partial_\beta \Phi \cdot \varepsilon_{\alpha\rho}
B^\rho + \eta_{\alpha\beta} \varepsilon^{\rho\sigma} (\partial_\rho B_\sigma
+ \frac{1}{2}\; B_\rho \times B_\sigma)
\cdot\Phi,
\end{array} \eqno(4.22)
$$
where the subscripts have the same meaning as explained earlier, and 
$$
\begin{array}{lcl}
X_{\alpha\beta}^{(N)} &=& 
\frac{1}{2} \partial_\alpha
\bigl [ \frac{1}{\xi}
(\partial_\rho A^\rho) \cdot A_\beta + E \cdot \varepsilon_{\beta\rho}
A^\rho \bigr ] 
+ \frac{1}{2} \partial_\beta
\bigl [ \frac{1}{\xi} (\partial_\rho A^\rho) \cdot A_\alpha + E \cdot 
\varepsilon_{\alpha\rho} A^\rho \bigr ] \nonumber\\
&-& \frac{i}{2}\; \eta_{\alpha\beta}\; \partial_\rho
\bigl [ \partial^\rho \bar C \cdot C + \bar C \cdot D^\rho C \bigr ].
\end{array} \eqno(4.23)
$$
It is clear from (4.22) that the symmetric energy momentum tensor 
mathematically corresponds to the $\bar\theta$-component of the local
(but composite) chiral superfields $Y_{\alpha\beta}^{(N)}$ and
$Z_{\alpha\beta}^{(N)}$. Here we substitute for the superfields, the
expansions in (3.35) and (3.38), which were derived after the application
of (dual) horizontality conditions. In fact, equation (4.22) captures
precisely the (anti-)commutators of (3.13) for the energy momentum
tensor $T_{\alpha\beta}^{(N)}$ in the language of chiral superfields.
As far as the geometry on the supermanifold is concerned, it is obvious
that the symmetric energy momentum tensor for the theory corresponds to
the translations of the composite superfields $Y_{\alpha\beta}^{(N)}$
and $Z_{\alpha\beta}^{(N)}$ along the $\bar\theta$-direction of the
supermanifold. This process of translation,
generated by the (co-)BRST charges,  corresponds to the (co-)BRST
transformations $s_{(d)b}$ on the ordinary 
(spacetime dependent) local (but composite) fields $L_{\alpha\beta}^{(1)}$
and $L_{\alpha\beta}^{(2)}$ of equation (3.13).\\

\noindent
{\bf 5 Conclusions}\\

\noindent
In the present investigation, we have 
elucidated the derivation of on-shell and off-shell
nilpotent (anti-)BRST and (anti-)co-BRST symmetries 
for the 2D (non-)Abelian gauge theories in the superfield formulation
by exploiting the (dual) horizontality conditions w.r.t. the super 
cohomological operators $\tilde d$ and $\tilde \delta$, defined on 
the $(2+2)$-dimensional supermanifold. For the 
derivation of the on-shell nilpotent symmetries, we have invoked  the
(anti-)chiral superfields which turn out to be quite handy and helpful. 
In fact, we have derived a mapping between the translations generators
$\partial/\partial\theta, \partial/ \partial\bar\theta$
and the internal nilpotent transformations of the on-shell variety
$s_{(a)b}, s_{(a)d}$ 
as well as the off-shell variety $\tilde s_{(a)b}, \tilde s_{(a)d}$ as 
$$
\begin{array}{lcl}
&&{\displaystyle \frac{\partial}{\partial \bar\theta}} \leftrightarrow s_{(d)b},
\qquad
{\displaystyle \frac{\partial}{\partial \theta}} \leftrightarrow s_{ab},
\qquad
{\displaystyle \frac{\partial}{\partial \theta}} \leftrightarrow s_{ad},
\nonumber\\
&&\tilde s_{(d)b} \leftrightarrow 
\mbox{ Lim}_{\theta, \bar\theta \rightarrow 0}\;
{\displaystyle \frac{\partial}{\partial \bar \theta}}\; \qquad
\tilde s_{ab} \leftrightarrow  
\mbox{ Lim}_{\theta, \bar\theta \rightarrow 0}\;
{\displaystyle \frac{\partial}{\partial \theta}}\; \qquad
\tilde s_{ad} \leftrightarrow  
\mbox{ Lim}_{\theta, \bar\theta \rightarrow 0}\;
{\displaystyle \frac{\partial}{\partial \theta}}.
\end{array}\eqno(5.1)
$$
This mapping enables us to provide the geometrical interpretation for the
nilpotent (anti-) BRST and (anti-)co-BRST charges (and transformations
they generate),  topological invariants (zero-forms), Lagrangian density
and symmetric energy momentum tensor for the 2D (non-)Abelian gauge theories. 
In the language of superfield
formulation, it turns out that the topological nature of the 2D free Abelian-
and self-interacting non-Abelian gauge theories is encoded in the form
of the Lagrangian density and symmetric energy momentum tensor which are found
to be the {\it total} derivatives w.r.t. Grassmannian variables. In view of
the above mapping, these physically and topologically interesting quantities
are equal to the sum of (co-)BRST anti-commutators. This property
establishes the fact that there are no energy excitations in the theory
because $<phys | \hat T^{00} | phys^{\prime}> = 0$ due to the fact that
$Q_{(d)b} |phys> = 0$. The nilpotency ($Q_{(a)b}^2 = Q_{(a)d}^2 = 0$)
of the (anti-)BRST and (anti-) co-BRST charges is also
encoded in the mapping (5.1) because $(\partial/ \partial \bar\theta)^2
 = 0, (\partial / \partial\theta)^2 = 0$. We have discussed in detail
the geometrical aspects of the above topologically interesting quantities in
the language of {\it translations} along some specific 
Grassmannian direction(s) of a four $(2+2)$-dimensional supermanifold.

One of the interesting features of our investigation is the observation
(and its proof) that the (dual) horizontality conditions on the (anti-)chiral
superfields lead to the derivation of the on-shell nilpotent (anti-)BRST
and (anti-)co-BRST symmetries that co-exist for the Lagrangian density of a
2D free Abelian gauge theory. The same does not happen in the case of
self-interacting 2D non-Abelian gauge theory. In fact, in some sense, the 
{\it chiral} superfield formulation provides an explanation for the 
non-existence of the on-shell nilpotent anti-BRST and anti-co-BRST symmetries 
for the Lagrangian density (3.1)
which respects the on-shell nilpotent (co-)BRST symmetries.
We have been able to establish the above fact by taking into account
the anti-chiral superfield and showing that no logically consistent nilpotent
symmetries emerge when we exploit the (dual) horizontality conditions.
Furthermore, we have shown that (anti-)chiral superfields can merge
consistently in the case of Abelian gauge theories and they 
shed some new light on the symmetries of the Lagrangian density and
symmetric energy momentum tensor. For instance, it can be emphasized that
the derivation of equations (4.5) and (4.16) from their superfield
versions (4.3) and (4.15) is a {\it completely new observation} which
was not seen, so far, in our earlier works [29,30,35]. This consistent 
merging of the (anti-)chiral superfields for the case of the 2D
non-Abelian gauge theory is impossible. This is the key reason behind the fact
that the anti-chiral superfields, on their own, do not lead to any logically
correct nilpotent symmetries even though we exploit the (dual) horizontality
conditions correctly w.r.t. the super cohomological
operators $\tilde d$ and $\tilde \delta$.
Thus, the {\it on-shell nilpotent}
anti-BRST and anti-co-BRST symmetries do not exist for the non-Abelian
gauge theory. In other words, the Lagrangian density for the 2D 
self-interacting non-Abelian gauge theory that respects the on-shell nilpotent
(co-)BRST symmetries does not respect the on-shell nilpotent anti-BRST
and anti-co-BRST symmetries.
For the derivation of the off-shell version of these
symmetries one has to invoke the most general superfield 
expansions as in (2.46).

In our earlier works [29-35], 
we were able to obtain the analogues of the de Rham
cohomology operators $(d,\delta,\Delta)$ in the language of
symmetry properties (and corresponding generators) for a given Lagrangian 
density of a gauge theory. In fact, in our present paper, we considered
the symmetries of the Lagrangian density for the 2D free- as well as
self-interacting (non-)Abelian gauge theories
and we demonstrated their existence in the (chiral) superfield formulation too. 
We found a two-to-one mapping:
$Q_{b(ad)} \rightarrow d, Q_{d(ab)} \rightarrow \delta, \{Q_{b}, Q_{d} \}
= \{Q_{ab}, Q_{ad} \} \rightarrow \Delta$ between the nilpotent
and conserved charges of the theory and the de Rham cohomological operators. 
The algebra of these charges (cf. eqn. (2.6)) with the ghost charge is 
such that (see, e.g., Ref. [35])
$$
\begin{array}{lcl}
i Q_{g} Q_{b(ad)} |\chi>_{n} &=& (n + 1)\; Q_{b(ad)} |\chi>_{n}, \nonumber\\
i Q_{g} Q_{d(ab)} |\chi>_{n} &=& (n - 1)\; Q_{d(ab)} |\chi>_{n}, \nonumber\\
i Q_{g} Q_{w} |\chi>_{n} &=& (n)\; Q_{w} |\chi>_{n}, 
\end{array} \eqno(5.2)
$$
where $|\chi>_{n}$ is an arbitrary state in the quantum Hilbert space with
the ghost number $n$ (i.e. $i Q_{g} |\chi>_{n} = n |\chi>_{n}$). Thus, the
analogue of the HDT (cf. (1.1)) can be easily defined in the quantum Hilbert 
space of states as given below [30,35] 
$$
\begin{array}{lcl}
|\chi>_{n} = |\omega>_{n} + Q_{b(ad)} |\theta>_{n-1} + Q_{d(ab)} |\phi>_{n+1},
\end{array} \eqno(5.3)
$$
where $|\omega>_{n}$ is the harmonic state (i.e. $Q_{b(ad)} |\omega>_{n} = 0,
Q_{d(ab)} |\omega>_{n} = 0$), $Q_{b} |\theta>_{n-1}$ is a BRST exact state
(which is equivalent to an anti-co-BRST exact state $Q_{ad} |\theta>_{n-1}$) 
and $Q_{d} |\phi>_{n+1}$ is a co-BRST exact state (which is equivalent to an
anti-BRST exact state $Q_{ab} |\phi>_{n+1}$). However, these considerations 
do not explain the reason 
behind the existence of the two-to-one mapping which we discussed above.
{\it It is primarily the geometrical superfield approach 
to BRST formalism that clarifies the 
existence of a two-to-one mapping between the conserved charges and the
cohomological operators}. In fact, the (dual) horizontality conditions w.r.t.
$\tilde d $ and $\tilde \delta$ imply that $\tilde d \rightarrow Q_{(a)b},
\tilde \delta \rightarrow Q_{(a)d}$. The ghost number consideration
(cf. (5.2)), however, clinches the issue and establishes the fact that there
is a two-to-one mapping between (anti-)BRST and (anti-)co-BRST
charges on the one hand and (co-)exterior derivatives of the
differential geometry on the other hand as: $ Q_{b(ad)} \rightarrow d,
Q_{d(ab)} \rightarrow \delta$. It will be noticed that we did not lay any 
emphasis on the (super) Laplacian operators $(\tilde \Delta)\Delta$ in
our present paper. This is because of the fact that the symmetry generated by
this operator is derivable from the nilpotent symmetries generated by
the (super) cohomological operators $(\tilde d)d$ and 
$(\tilde \delta)\delta$. Furthermore, geometrically, the
(super) Laplacian operators $(\tilde \Delta)\Delta$ do not lead to
any impressive results. For the 2D free Abelian gauge theory, this exercise
with $(\tilde \Delta)\Delta$ was performed (together with the analogue of 
the horizontality condition w.r.t. these operators) in [38]. This attempt,
with the help of certain specific discrete symmetries of the theory, led
to the derivation of a bosonic symmetry only for 
the Lagrangian density (2.1) {\it but not for (2.3)}.

We concentrated, in our present work, only on the 2D free Abelian- and 
self-interacting non-Abelian (one-form) gauge theories. However, our hope
is to extend our understanding and insight of 2D one-form gauge theories
to the physical 4D two-form gauge theories where we have been able to show the 
existence of (anti-)BRST and (anti-)co-BRST symmetries in the Lagrangian
formulation [34]. The study of the geometrical aspects of the 4D  two-form
gauge theories in the superfield formulation (together with the
understanding of the new symmetries) might turn out to be useful 
in the context of (super)string theories where one considers
a non-trivial (spacetime dependent) metric. The
analogue of the new symmetries might shed some new light on the proof of 
renormalizability of the interacting two-form gauge theories where matter
fields are coupled to the two-form potentials. Another direction that 
can be pursued, following
the approach adopted in [51], is to try for the superfield formulation
of the Batalin-Vilkovsky formalism applied to the one-form and two-form
gauge theories with an extended set of BRST operators. This extended set will
include, in addition to the nilpotent (anti-)BRST charges [51],
the nilpotent (anti-)co-BRST charges and a bosonic
conserved charge, too. The discussion of  the
new dual BRST symmetries in the context of three 
dimensional BF system (see, e.g., Ref. [52] for details) and its 
application to the study of topological 
properties of this system, etc., are yet another 
few novel directions that can be taken up
for considerations as far as  our future endeavours connected with the
superfield formalism are concerned. We have enumerated here
some of the issues and pointed out a few directions
that are under investigation and our results would be 
reported in our future publications [53].

\end{document}